\documentclass[twocolumn,amsmath,amssymb,aps,prr,longbibliography]{revtex4-1}
\usepackage{graphicx}
\begin{document}

\title{Chiral Edge Currents For ac Driven Skyrmions In Confined Pinning Geometries}
 
\author{C. Reichhardt and C.J. Olson Reichhardt}
\affiliation{Theoretical Division and Center for Nonlinear Studies,
Los Alamos National Laboratory, Los Alamos, New Mexico 87545, USA}

\date{\today}

\begin{abstract}
We show that ac driven skyrmion lattices in a weak pinning channel confined by 
regions of strong pinning exhibit edge transport
carried by skipping orbits 
while skyrmions in the bulk
of the channel undergo localized orbits with no net transport.
The magnitude of 
the edge currents can
be controlled by varying the amplitude and frequency
of the ac drive or by changing the ratio of the Magnus force to the damping term. 
We identify a localized phase in which the orbits are small
and edge transport is absent,
an edge transport regime,
and a fluctuating regime that appears when the ac drive is strong
enough to dynamically disorder the skyrmion 
lattice.
We also find that in some cases,
multiple rows
of skyrmions participate in the
transport due to a drag effect
from the skyrmion-skyrmion interactions. 
The edge currents are robust for
finite disorder and should be a general 
feature of skyrmions interacting with confined geometries or
inhomogeneous disorder under an ac drive.
We show that similar
effects can occur for skyrmion lattices at
interfaces or along domain boundaries for multiple coexisting
skyrmion species.
The edge current effect
provides a new method to control skyrmion motion,
and we discuss the connection of these results with
recent studies
on the emergence of edge currents in chiral active matter systems
and gyroscopic metamaterials. 
\end{abstract}

\maketitle

\vskip 2pc

\section{Introduction}

A paradigmatic example of
a system that exhibits edge states or edge currents in confinement is
electrons in a magnetic field undergoing cyclotron motion.
Here, bulk electrons follow closed circular orbits 
while charges near the boundaries enter
skipping orbits,
and the direction of the resulting
current is determined by the 
chirality of the cyclotron motion
\cite{Teller31,Halperin82,Beenakker89,Muller92,Reijniers02}.
Edge currents also arise in cold atom systems \cite{Stuhl15}, chiral active 
matter spinners \cite{vanZuiden16,Han17,Dasbiswas18,Reichhardt19},
coupled gyroscopes \cite{Nash15,Susstrunk15,Mitchell18,Mitchell18a},
and colloids
placed on
periodic magnetic substrates to create colloidal topological insulators
\cite{Loehr16,Loehr18} in analogy to electronic
topological insulators \cite{Hasan10}.
In all of these systems, there is some form of periodic orbit 
with a particular chirality as well as some type of boundary or interface.
Edge currents can also arise for
magnetic skyrmions,
which have dynamics that are similar in many ways to those
of electrons in a magnetic field.
Skyrmions can exhibit
complex cyclotron orbits with a fixed chirality,
and
in certain sample geometries,
the skyrmions interact with some form of boundary or interface
\cite{Nagaosa13,EverschorSitte18}.   

Skyrmions are particle like
textures that were originally proposed to appear in magnets
in 1989 \cite{Bogdanov89}.
Magnetic skyrmion lattices were experimentally
observed in 2009 
using neutron scattering \cite{Muhlbauer09}
and were directly imaged with Lorentz microscopy \cite{Yu10}. Since  
these initial observations,
an increasing variety of materials have been identified which
support skyrmions, including systems in which
skyrmions are stable at room temperature
\cite{Jiang15,Tokunaga15,Woo16,Soumyanarayanan17,Jiang17}.
When skyrmions are set into motion
by an applied current
\cite{Schulz12,Yu12,Iwasaki13,Lin13a,Liang15,Legrand17,Tolley18}, 
they
exhibit depinning and sliding phases similar to those
found for vortices in type-II superconductors
and Wigner crystals \cite{Reichhardt17}.
Due to their size scale and the fact that they can be moved easily with a current,
skyrmions are also
candidates for possible memory and computing
applications \cite{Fert13,Tomasello14,Fert17,Zazvorka19},  
and the understanding
of skyrmion dynamics on the individual and collective level
will be integral to the creation of such devices. Although
skyrmions have many similarities to other particle-based
systems that exhibit depinning, they also have 
several distinct properties, the most prominent of which is the
domination of skyrmion dynamics by the Magnus force.
Skyrmion motion has
many similarities to the dynamics of electrons in a magnetic field;
however, the skyrmions can also experience significant
damping, and the ratio of the
Magnus force to the damping force
depends on the material parameters and
can produce
different dynamical effects \cite{Nagaosa13,EverschorSitte18,Jiang17,Schulz12,Lin13}.
The Magnus force generates a skyrmion velocity component
that is perpendicular to the net external force acting on the skyrmion.
One consequence of a finite Magnus term is that 
the skyrmions move at an angle with respect to an applied 
drive that is known as the skyrmion Hall angle
\cite{Nagaosa13,EverschorSitte18,Muhlbauer09}.
In the absence of quenched disorder or 
pinning, the skyrmion Hall angle is constant;
however, when
quenched disorder is present,
the skyrmion Hall angle becomes drive or velocity dependent, starting at
a value of nearly zero
just above the depinning threshold 
and approaching the disorder-free limit at higher drives
\cite{Reichhardt15,Reichhardt16,Kim17,Jiang17a,Litzius17,Woo18,Juge19,Zeissler19}.
The Magnus force also causes
the skyrmions
to exhibit cyclotron or spiraling motion when they are in a confining potential or  
interacting with a pinning site
\cite{Reichhardt15,Liu13,Lin13b,Muller15,Buttner15,Navau16,GonzalezGomez19,Liu19,Menezes19}. 
Skyrmions can perform circular orbits
under biharmonic drives \cite{Chen19},
oscillating fields \cite{Moon16,Yuan19},
and in certain types of driven bilayer systems \cite{Ritzmann18}. 
In numerical studies of skyrmions with one-dimensional (1D)
periodic and asymmetric substrate arrays, skyrmions
under ac, dc, and combined ac and dc drives
exhibited complex periodic closed and running orbits
\cite{Reichhardt15b,Reichhardt16a,Reichhardt17a}. 

Since skyrmions can undergo chiral motion when interacting with disorder 
or subjected to a drive,
it is natural 
to look for edge currents in the presence of
a confining geometry when the skyrmions
are driven or perturbed in some way.
In this paper,
we examine skyrmion lattices in a system
containing a pin-free channel
surrounded by strong pinning, with an initially uniform distribution of
skyrmions in the entire sample.
Under an ac drive,
the skyrmions in the pin-free channel undergo periodic motion.
If there were no confinement,
the periodic orbits would take the form of 1D paths oriented at an angle to the
applied drive; however, when confinement is present,
the skyrmions near the edges
of the pin-free channel
follow circular or elliptical orbits
with a chirality that is
determined by the sign of the Magnus force.
These orbits interact with the edge potential
created by the pinned skyrmions and become
skipping orbits that generate
edge currents moving to the right or left, depending on the side of the
pin-free channel at which they appear.
The skyrmions in the bulk of the pin-free channel remain localized.
We find slip phenomena for the skyrmions participating in the edge transport,
and the effectiveness of the edge currents depends on the ratio of the Magnus force
to the damping force as well as on the amplitude and frequency of the ac drive and on
the skyrmion density.

We identify three regimes of behavior.
For ac drives with high frequency or low amplitude, we observe a localized
phase in which no edge transport occurs
but the orbits at the edges of the pin-free channel
are much more circular than
the orbits in the bulk of the pin-free channel.
We also find a regime of 
edge transport in which
the skyrmions in the pin-free channel form a lattice,
as well as a liquid regime in which
the ac drive is large enough 
to melt the skyrmion lattice dynamically.
Within the liquid regime, near the transition to the lattice regime,
edge transport can still appear in which
skyrmions at the edge of the pin-free channel
can be transported over some distance before exchanging with skyrmions in the bulk
of the pin-free channel;
however, deeper within the liquid regime,
the edge transport is lost. 
In the limit of zero Magnus force,
there is no edge transport and the orbits are 1D throughout
the sample.  
The edge transport we  observe is a collective effect, and it does
not appear
in the low density limit.

We also find that the
width and type of orbit along the edge
of the pin-free channel depend on the direction 
of the applied ac drive.
When the drive is parallel to the channel,
pronounced edge transport appears, while a drive that
is perpendicular to the channel
generally produces less edge transport since the
orbits become 1D and are oriented parallel to the edge of the channel.
In regimes where
strong edge transport occurs, the moving skyrmions along the edge
can
drag the adjacent rows of skyrmions,
causing these rows to exhibit a drift of reduced magnitude.
These effects are robust for varying amounts of disorder, and in some
cases,
a disorder-free system that shows no edge transport can develop
edge currents when disorder is introduced.
In the low density regime the edge transport is generally weak or absent, 
but we find that commensurate-incommensurate
effects between 1D rows of skyrmions  can create
currents flowing in the bulk of the pin-free channel instead of along its edges.
Such effects
occur when the edge row and an adjacent row
contain different numbers of skyrmions, producing
a skyrmion gear motion.
Finally,
we
show that edge currents should be a general feature that appears at
any type of skyrmion interface,
such as along the domain wall separating two different species of
skyrmions.
These effects should arise for confining geometries under ac drives
in samples with inhomogeneous pinning,
edge roughness, grain boundaries, or twin boundaries.
Our results
provide a new method for
transporting skyrmions
that avoids the skyrmion Hall effect found for dc driven skyrmions.

The paper is organized as follows.
In Section II we discuss the system and the simulation method. 
In section III,
we describe the conditions
under which edge currents arise and
demonstrate the different dynamical regimes.
Section IV shows the effect of
applying the ac driving along different directions,
while in
Section V we examine the role of disorder.
In Section VI we vary the skyrmion density and explore
skyrmion pumping produced by
commensurate-incommensurate effects,
while in section VII we discuss our results,
describe geometries
in which the edge currents could arise,
and show that
edge currents can
occur even for pin-free systems
at an interface between different skyrmion species.  
In Section VIII we summarize our results. 

\section{Simulation}

\begin{figure}
\includegraphics[width=3.5in]{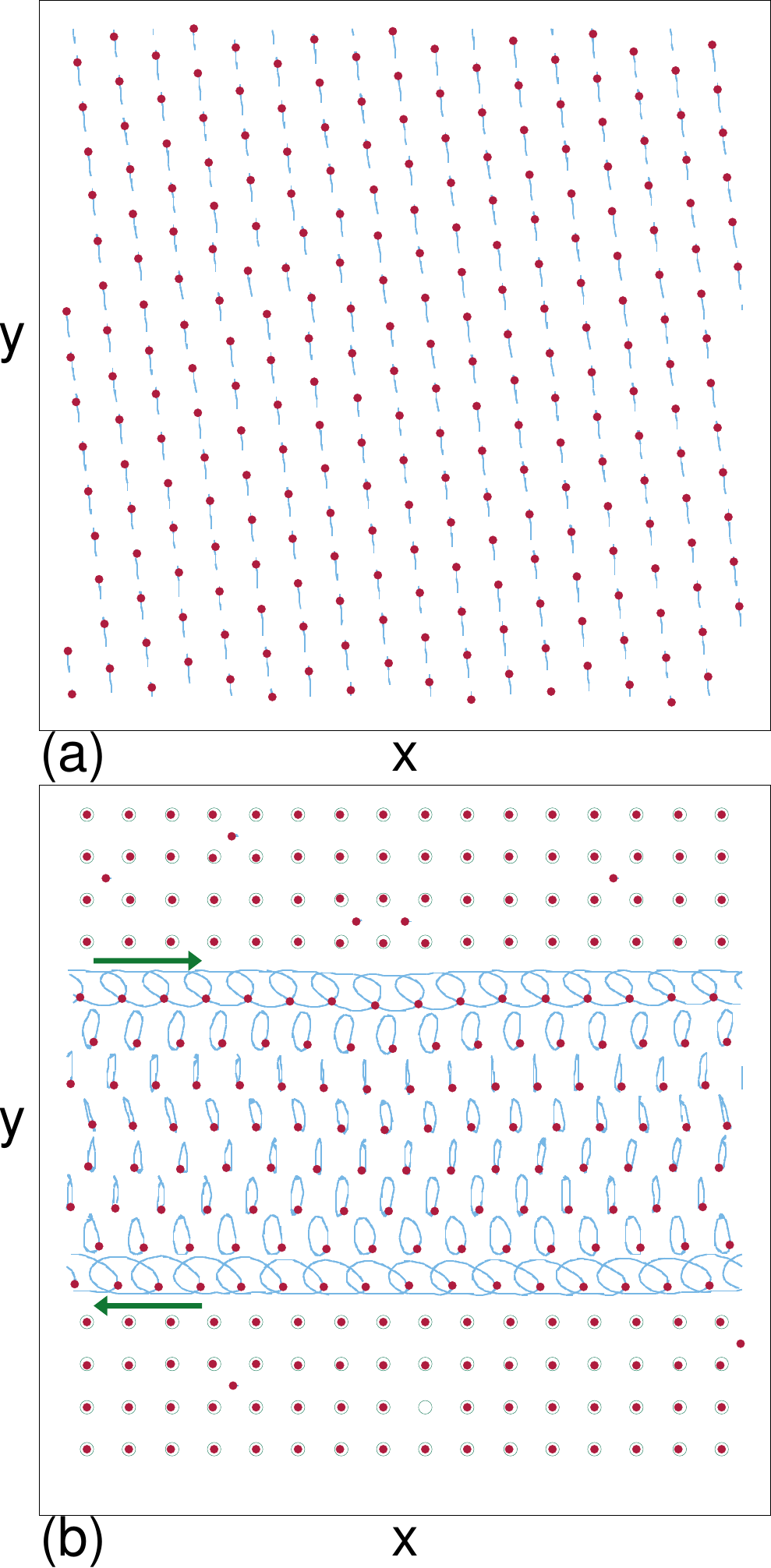}
\caption{(a) Skyrmion positions (dots) and trajectories (lines) for a
  system without any pinning at a skyrmion density of $n_{sk}=0.2$
  and a matching density of $n_{\phi}=0.2$
for $\alpha_m/\alpha_d=10.0$ under
an $x$-direction ac drive with amplitude $A=0.075$ and
frequency $\omega=3.75 \times 10^{-5}$.
The skyrmions
form a triangular lattice and move at an angle of $\theta = 84^\circ$
with respect to the driving direction. 
(b)  The same for a sample
in which half of the system is filled with a square
array of pinning sites
(open circles).
The skyrmions in the pin-free channel
form a triangular lattice and execute closed counterclockwise orbits,
while the
skyrmions in the pinned region remain immobile.
Edge transport occurs along the boundaries of the pin-free channel,
as indicated by the arrows.
}
\label{fig:1}
\end{figure}

We consider a two-dimensional (2D)
system of size $L \times L$
with periodic boundary conditions in the $x$ and $y$-directions.
Half of the sample contains a square array of pinning sites, and the other
half of the sample consists of a pin-free channel aligned with the $x$ direction.
The system contains
$N_{sk}$ skyrmions and $N_{p}$ pinning sites.
The skyrmion density is $n_{sk}=N_{sk}/L^2$ and the matching density at which
the number of skyrmions would equal the number of pins in a sample with
uniform pinning is $n_{\phi}=2N_p/L^2$.
The initial positions of the skyrmions are
obtained using simulated annealing.
The skyrmion-skyrmion interactions are
repulsive,
and in the absence of pinning, the skyrmions
form a uniform triangular
lattice.
After annealing,
we apply an ac drive
parallel or perpendicular to the pin-free channel.
In Fig.~\ref{fig:1}(a) we illustrate the skyrmion locations
and trajectories in a pin-free system under an
ac drive applied along the $x$-direction.
Here the skyrmions form a triangular lattice and execute
1D orbits oriented at an 
angle with respect to the drive direction
given by the skyrmion Hall angle $\theta^{\rm int}_{sk}$. 
In Fig.~\ref{fig:1}(b),
half of the sample contains a square pinning lattice and a pin-free channel
is oriented along the $x$ direction.
Here, when an ac drive is applied,
the skyrmions in the pinning sites remain immobile and produce
a confining potential for the skyrmions in the
pin-free channel.
Within the pin-free channel, the skyrmions
form a triangular lattice but
no longer undergo the strictly
1D motion
of Fig.~\ref{fig:1}(a).
Instead, the skyrmions follow
circular or elliptical orbits, with a continuously translating or edge current
appearing along the edges of the channel,
as highlighted by the arrows.
The skyrmions in the bulk of the pin-free channel perform
localized orbits.
In this case, the skyrmion
orbits are counterclockwise,
so the skyrmions on the bottom edge
of the pin-free channel are moving in the $-x$
direction while those on the top edge of the channel
are translating in the $+x$ direction.

We model the skyrmions using a particle based approach for
skyrmions interacting with pinning as
employed previously \cite{Lin13,Reichhardt15,Reichhardt15b,Reichhardt16a,Reichhardt17a,Brown19}. 
The motion of skyrmion $i$ is
governed by the following equation of motion:    
\begin{equation} 
\alpha_d {\bf v}_{i} + \alpha_m {\hat z} \times {\bf v}_{i} =
{\bf F}^{ss}_{i} + {\bf F}^{sp}_{i}  + {\bf F}^{ac}_{i} .
\end{equation}
The first term on the right is the repulsive skyrmion-skyrmion
interaction force 
${\bf F}_{i} = \sum^{N}_{j\neq i}K_{1}(r_{ij}){\hat {\bf r}_{ij}}$,
where $r_{ij} = |{\bf r}_{i} - {\bf r}_{j}|$ is the distance between
skyrmions $i$ and $j$ and
$K_{1}(r)$ is the modified Bessel function which
decreases exponentially at large $r$.
The pinning force is given by ${\bf F}_i^{sp}$ and the
pinning sites are modeled 
as parabolic traps of maximum strength $F_{p}$
and radius $r_{p}$.
In this work we set
$F_{p}$ sufficiently large
that skyrmions
within the pinning sites
remain immobile for all the parameters we consider.
Additionally, $r_p$ is small enough that each pinning site
captures at most one skyrmion.
We focus on the regime in which there are twice as many skyrmions
as pinning sites,
so that the pin-free and pinned regions
contain equal numbers of skyrmions.

An ac driving force ${\bf F}^{ac}$ is applied
in either the $x$ or $y$ direction,
${\bf F}^{ac} = A\sin(\omega t)({\bf \hat x},{\bf \hat y})$.
The term $\alpha_{d}$ is the damping constant
which aligns the velocities in the direction of the net force while
$\alpha_{m}$ is the coefficient of the Magnus term 
which aligns the velocity  perpendicular to the net force. 
The dynamics can be characterized by the ratio $\alpha_{m}/\alpha_{d}$, 
and in the absence of pinning the skyrmions
move at an angle with respect to a dc drive 
known the intrinsic skyrmion Hall angle
$\theta^{\rm int}_{sk} = \arctan(\alpha_{m}/\alpha_{d})$.
When $\alpha_{m}= 0$, the
system is in the overdamped  limit. 
The initial positions of the skyrmions are
obtained by starting from a high
temperature liquid state and cooling
to $T = 0.0$ in order to obtain an
overall skyrmion density that is roughly constant in both the
pinned and unpinned regions.   

\section{Edge Transport and Dynamical Regimes}

\begin{figure}
\includegraphics[width=3.5in]{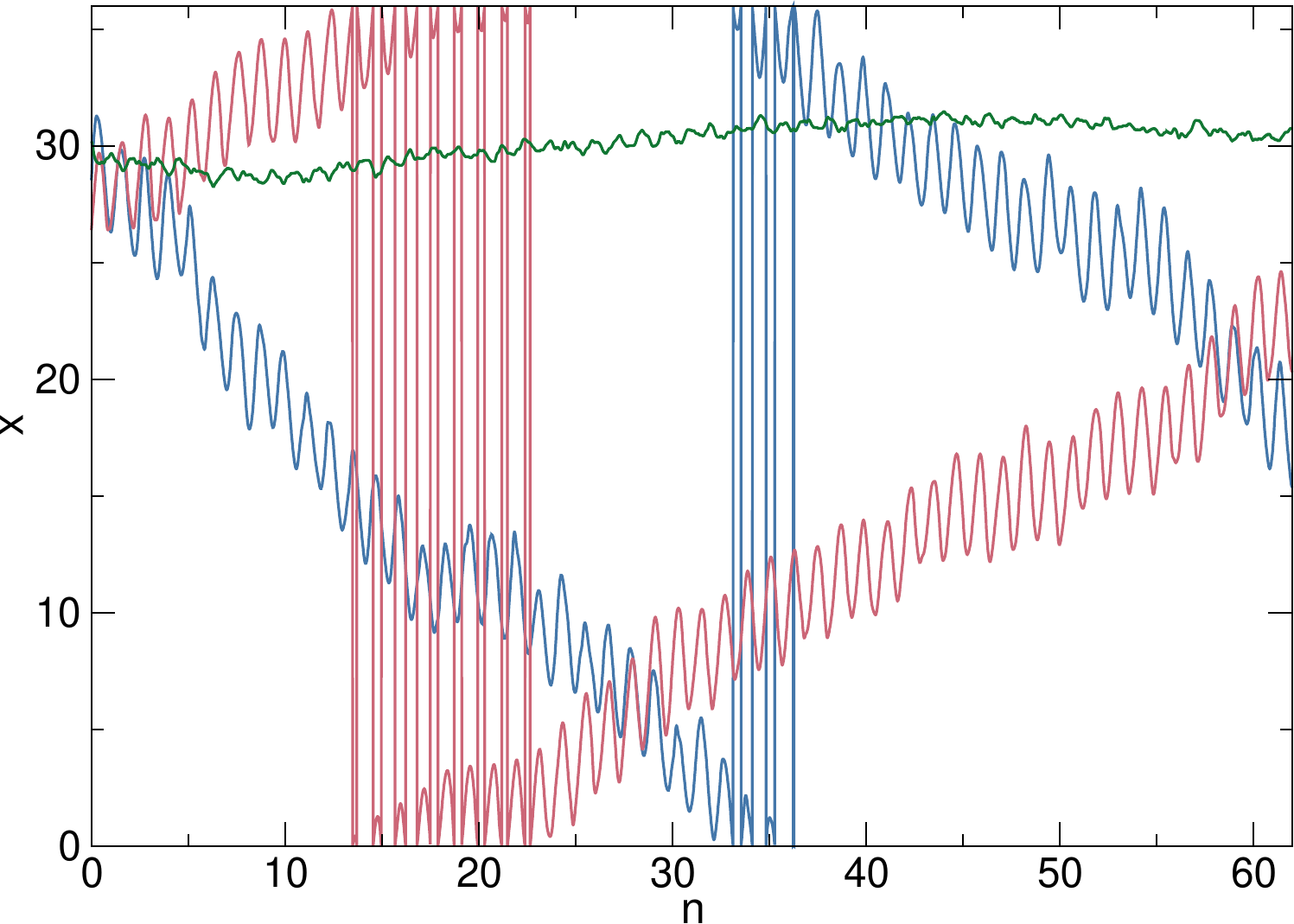}
\caption{ $x$ position vs ac cycle number $n$
  for individual skyrmions at the bottom edge of the pin-free channel (blue),
  the top edge of the pin-free channel (red),
  and in the center of the pin-free channel (green)
  in a sample with
  $x$ direction driving,
  $\alpha_{m}/\alpha_{d} = 10$,
  $A = 0.075$, $\omega = 3.75 \times 10^{-5}$,
  skyrmion density $n_{sk} = 0.4$, and matching density $n_{\phi}=0.4$.
  The skyrmion on the top edge moves in the positive
  $x$ direction,
  the skyrmion on the bottom edge
  moves in the negative $x$ direction,
  and the skyrmion in the center remains localized.
  The trajectories of these three skyrmions are illustrated in Fig.~\ref{fig:3}(b).
  The jumps between $x=0$ and $x=36$ occur when the skyrmion crosses the
  periodic boundary.
  }
\label{fig:2}
\end{figure}

\begin{figure}
\includegraphics[width=3.5in]{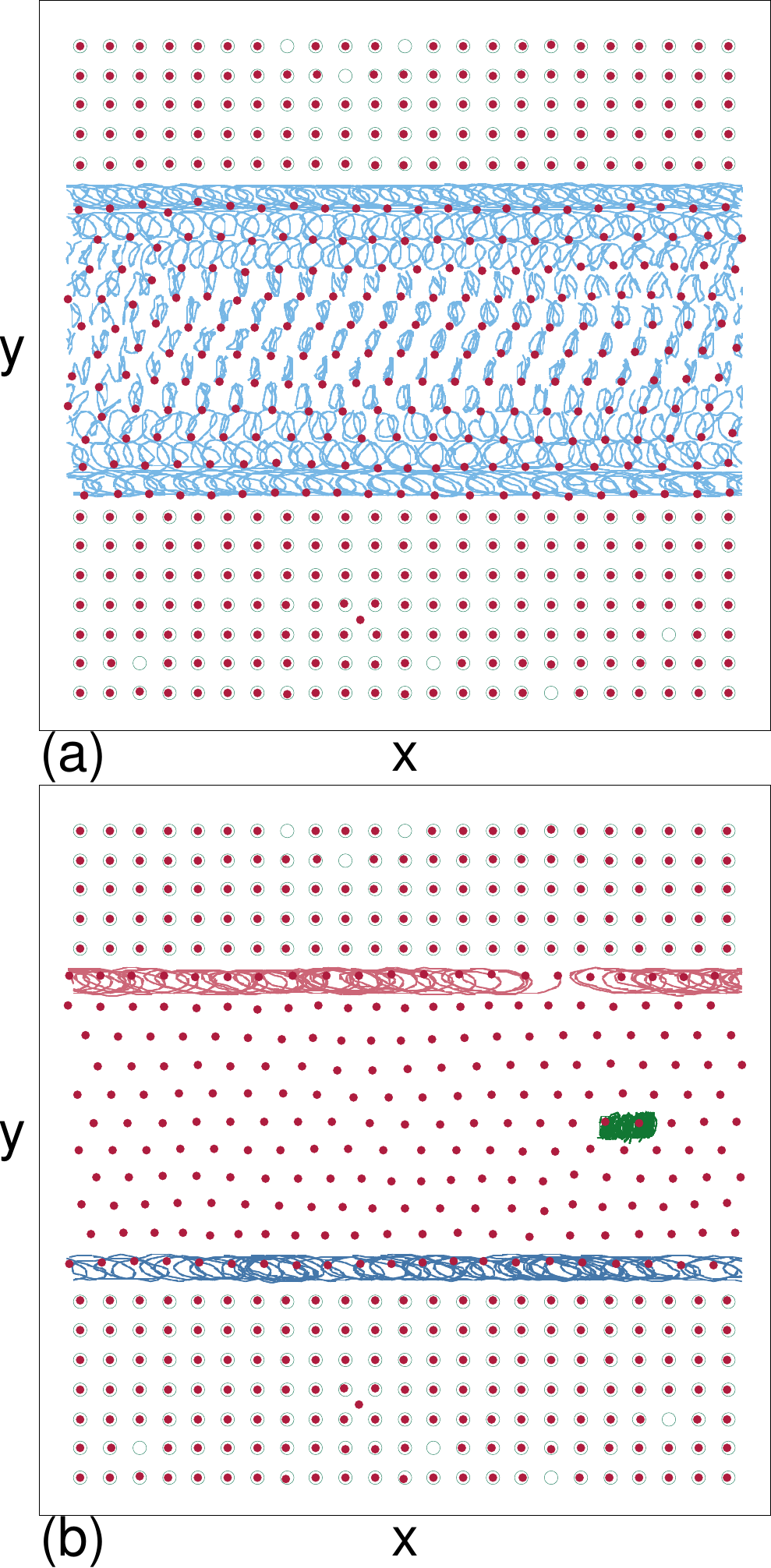}
\caption{Skyrmion positions (dots), all skyrmion trajectories (lines),
  and pinning site locations (open circles) for the system in 
  Fig.~\ref{fig:2}
  with $x$ direction ac driving illustrating the edge transport.
  Here $\alpha_m/\alpha_d=10$, $A=0.075$,
  $\omega=3.75 \times 10^{-5}$, and $n_{sk}=n_{\phi}=0.4$.
  (b) Image of the same system with the trajectories of only the three
  skyrmions shown in Fig.~\ref{fig:2} plotted.
  Red: skyrmion at the top of pin-free channel; green: skyrmion in
  the center of pin-free
  channel; blue: skyrmion at the bottom of the pin-free channel.}
\label{fig:3}
\end{figure}

We
first illustrate that skyrmions
translate along the edges
of the pin-free region.
In Fig.~\ref{fig:2} we plot the
$x$ position versus ac cycle number $n$
for individual skyrmions
on the bottom edge,
top edge,
and center of the pin-free channel
in a system with $\alpha_{m}/\alpha_{d} = 15$, 
$n_{sk} = 0.4$, $n_{\phi}=0.4$, $A = 0.075$, and $\omega = 3.75 \times 10^{-5}$.
All three skyrmions start at an $x$ position near $x=29$.
In Fig.~\ref{fig:3}(a) we illustrate
the skyrmion positions, pinning site locations, and trajectories for
all skyrmions
from the system in Fig.~\ref{fig:2},
while
in Fig.~\ref{fig:3}(b) we
trace the trajectories of only the three skyrmions
shown in Fig.~\ref{fig:2}.
The skyrmion on the bottom edge of the pin-free channel
moves in the $-x$ direction, undergoing a displacement of
$\delta x=-45$ during the course of 70 ac cycles, as shown
in Fig.~\ref{fig:2}.
The skyrmion on the top edge of the pin-free channel
moves a distance of $\delta x=30$ in the positive $x$ direction
during the same time period, and
the skyrmion at the center of the channel has no net displacement.
The skyrmions that are undergoing a net translation do not exhibit
completely periodic motion but occasionally become localized for
a period of time when phase slips occur,
producing the
disorder in the orbits
found in Fig.~\ref{fig:3}(b).
Since the lattice
constant of the pinning sites is $a = 1.56$,
if the skyrmion lattice was perfectly ordered and the
skyrmion orbit perfectly matched the pinning lattice size scale to produce 
an ideal locking between the cyclic orbit and
the periodic drive,
a total displacement of $|\delta x_{\rm ideal}|=110$ would occur during the course
of $n=70$ cycles.
The size of the elliptical orbit on the edges of the pin-free channel
depends on the frequency and amplitude of the ac drive
as well as on the value of
$\alpha_{m}/\alpha_{d}$, so the efficiency
$|\delta x|/|\delta x_{\rm ideal}|$
of the
transport current
depends on these parameters.

\begin{figure}
\includegraphics[width=3.5in]{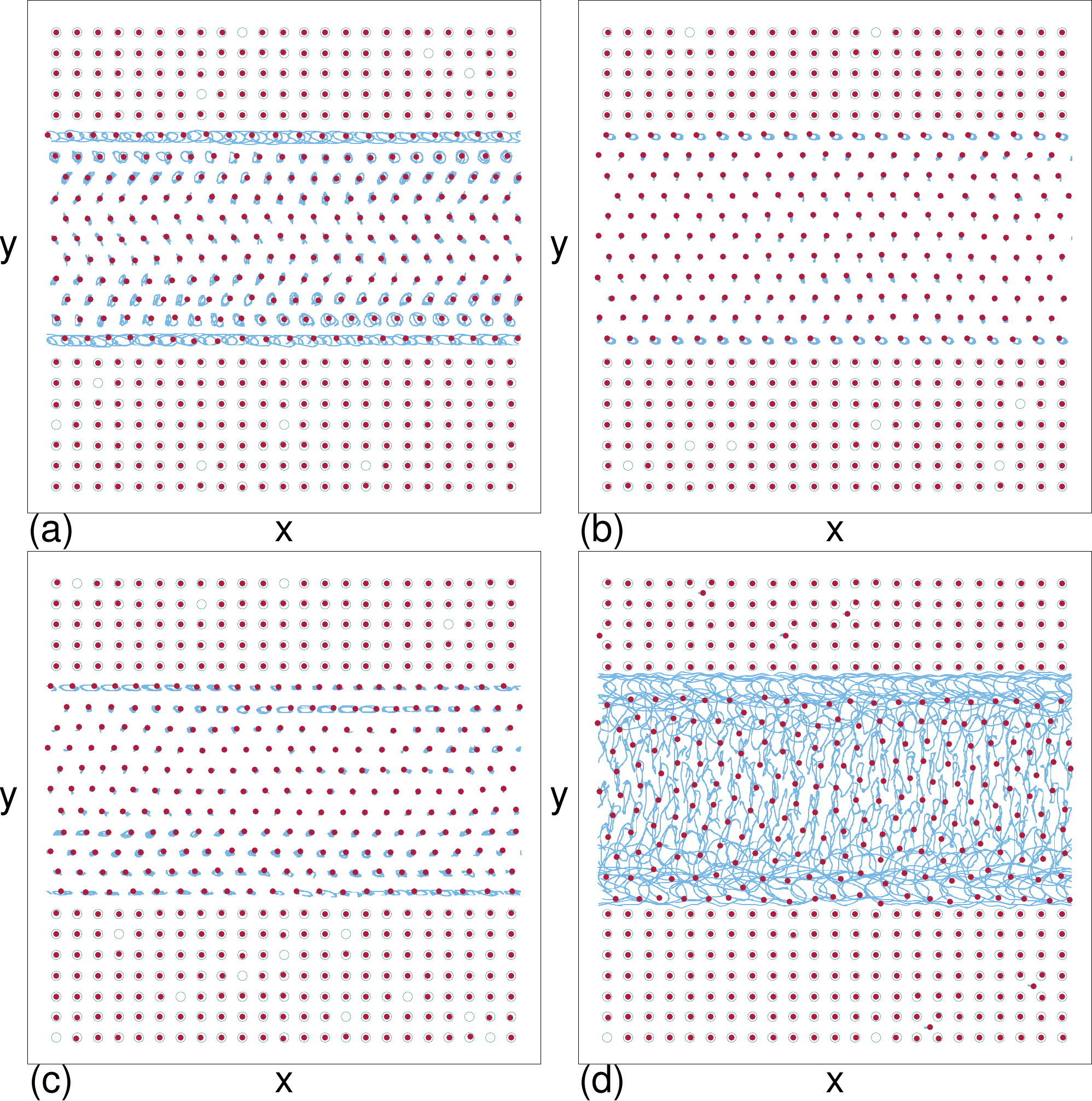}
\caption{Skyrmion positions (dots), trajectories (lines), and pinning site
  locations (open circles)
  for the system in Fig.~\ref{fig:2}
  at different $x$ direction ac drive frequencies and amplitudes,
  where $\alpha_m/\alpha_d=10$ and $n_{sk}=n_{\phi}=0.4$.
  (a) For $\omega=6 \times 10^{-5}$ and $A=0.075$,
  the orbits are smaller but edge transport persists.
  (b) At $\omega = 1 \times 10^{-4}$ and $A=0.075$,
  the orbits are small enough that
  no edge transport occurs. 
  (c) At $\omega=3.75 \times 10^{-5}$ and $A = 0.025$,
  there is also no edge transport due to the small size of the orbits.
  (d) At $\omega=3.75 \times 10^{-5}$ and $A  = 0.15$,
  the system is in a disordered or fluctuating state. }
\label{fig:4}
\end{figure}

In Fig.~\ref{fig:4} we
plot the skyrmion trajectories for the system from Fig.~\ref{fig:3}
at different values of the ac drive amplitude $A$ and frequency $\omega$.
Figure~\ref{fig:4}(a,b) shows
systems with $A = 0.075$ at higher frequencies of 
$\omega = 6 \times 10^{-5}$
and $\omega = 1 \times 10^{-4}$, respectively.
Edge currents are still present
when $\omega = 6 \times 10^{-5}$ in Fig.~\ref{fig:4}(a),
but in Fig.~\ref{fig:4}(b)
at $\omega = 1 \times 10^{-4}$,
the edge transport is lost, with
skyrmions at the edge of the pin-free channel executing
circular orbits
and skyrmions
in the bulk of the pin-free channel following
1D trajectories.
In general, as the ac drive frequency increases,
the size of the
skyrmion
orbits shrinks,
but
edge transport only occurs when the orbits are large enough
that they either overlap with each other or are wider than
the period of the
confining potential. 
Figure~\ref{fig:4}(c) shows
a sample with $\omega = 3.75 \times 10^{-5}$ at a smaller $A = 0.025$, where the orbits 
are small enough that the edge currents are lost.
In Fig.~\ref{fig:4}(d) at $\omega=3.75 \times 10^{-5}$ and 
$A = 0.15$,
the orbits 
are large enough that the skyrmions in the pin-free channel
become strongly disordered and form a dynamical
liquid or fluctuating state.
In this fluctuating regime, it is still possible for edge transport to occur
when skyrmions near the edges of the channel
translate for some distance before exchanging with
a skyrmion in the bulk of the pin-free channel.
Thus, we
observe both
a crystal with edge transport and a liquid with edge transport.
For larger values of $A$, the skyrmions
become more disordered and the edge transport is strongly reduced or absent,
and when $A$ is sufficiently large, the skyrmions at the pinning sites begin to depin
and the distinction between skyrmions in the
pin-free channel and those in the pinned region
is destroyed.

\begin{figure}
\includegraphics[width=3.5in]{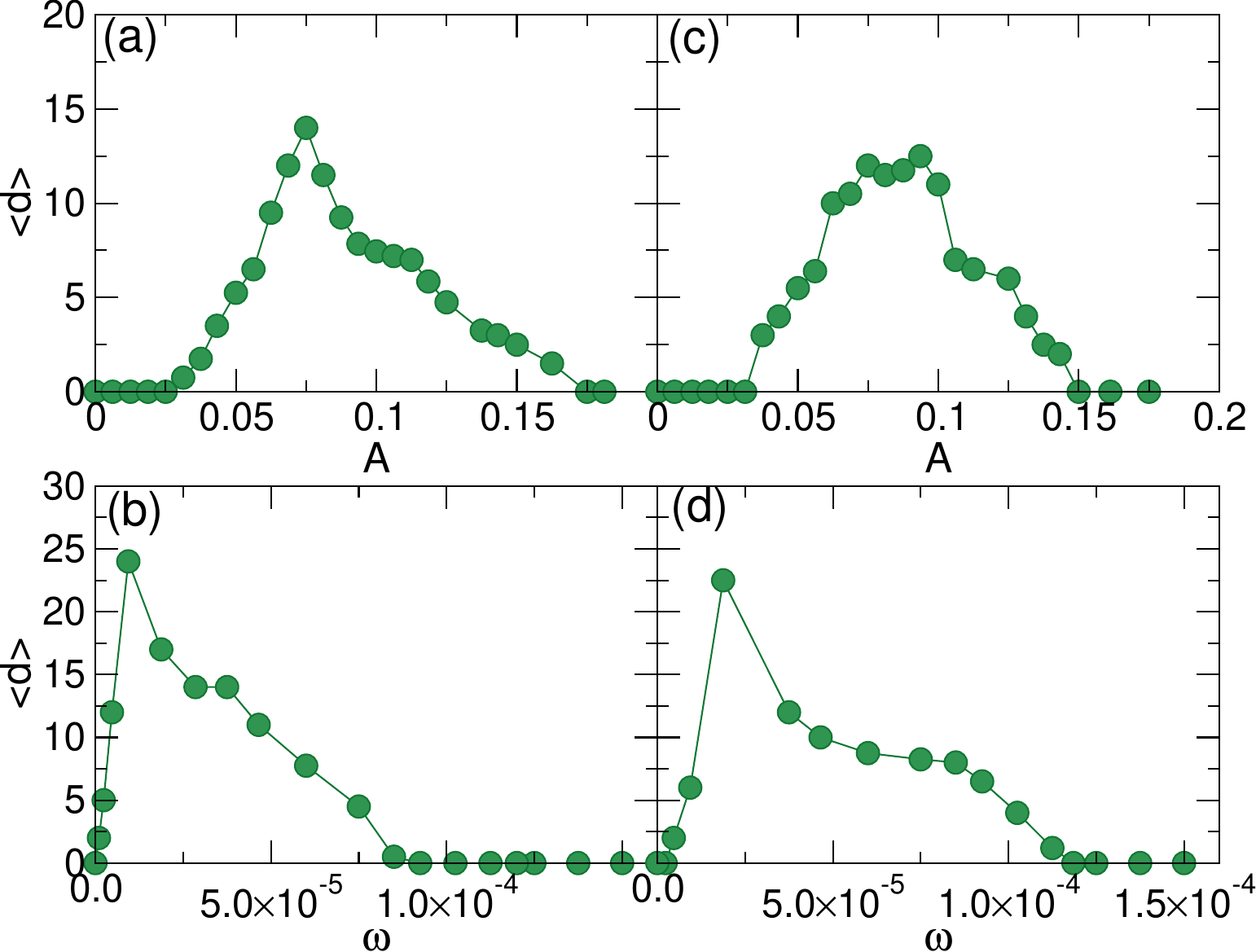}
\caption{The average drift distance $\langle d\rangle$ for the edge skyrmions
during 20 ac drive cycles
for the system in Fig.~\ref{fig:4}(a) with
$x$ direction driving and $n_{sk}=n_{\phi}=0.4$.
(a) $\langle d\rangle$ vs $A$ for
a sample with $\alpha_{m}/\alpha_{d} = 15$
at $\omega=3.75 \times 10^{-5}$.
(b) $\langle d\rangle$ vs $\omega$ for the system in
(a) at $A = 0.075$. 
(c) $\langle d\rangle$ vs $A$ for
a sample with $\alpha_{m}/\alpha_{d} = 5.0$
at $\omega = 3.75 \times 10^{-5}$.
(d)  $\langle d\rangle$ vs $\omega$ for the system in (c) 
at $A = 0.075$.}
\label{fig:5}
\end{figure}

In Fig.~\ref{fig:5} we plot the average
drift distance
$\langle d\rangle = N^{-1}_{\rm edge}\sum_{i=1}^{N_{\rm edge}} |\delta x_i|$ for
the $N_{\rm edge}$ skyrmions on the edges of the pin-free channel during a time
period of $n=20$ ac drive cycles.
Figure~\ref{fig:5}(a,b) shows
$\langle d\rangle$ versus $A$ and $\omega$, respectively,
for the system in Fig.~\ref{fig:4}
at $\alpha_{m}/\alpha_{d} = 15$.
In both cases there is an optimum drive parameter that maximizes the
edge transport,
while for low $A$ or high $\omega$,
the edge skyrmions are localized.
In Fig.~\ref{fig:5}(a), when $A > 0.115$ the skyrmion lattice
starts to become disordered  and forms a liquid state. 
Figure~\ref{fig:5}(b) indicates that the edge transport is
more efficient
at lower frequencies since fewer phase slips occur;
however, the orbits also become increasingly one-dimensional as $\omega$
decreases, so at small values of $\omega$ the edge transport is reduced.
In Fig.~\ref{fig:5}(c,d) we plot $\langle d\rangle$ versus $A$ and
$\omega$
for a system with 
$\alpha_{m}/\alpha_{d} = 5.0$,
where similar behavior appears.

\begin{figure}
\includegraphics[width=3.5in]{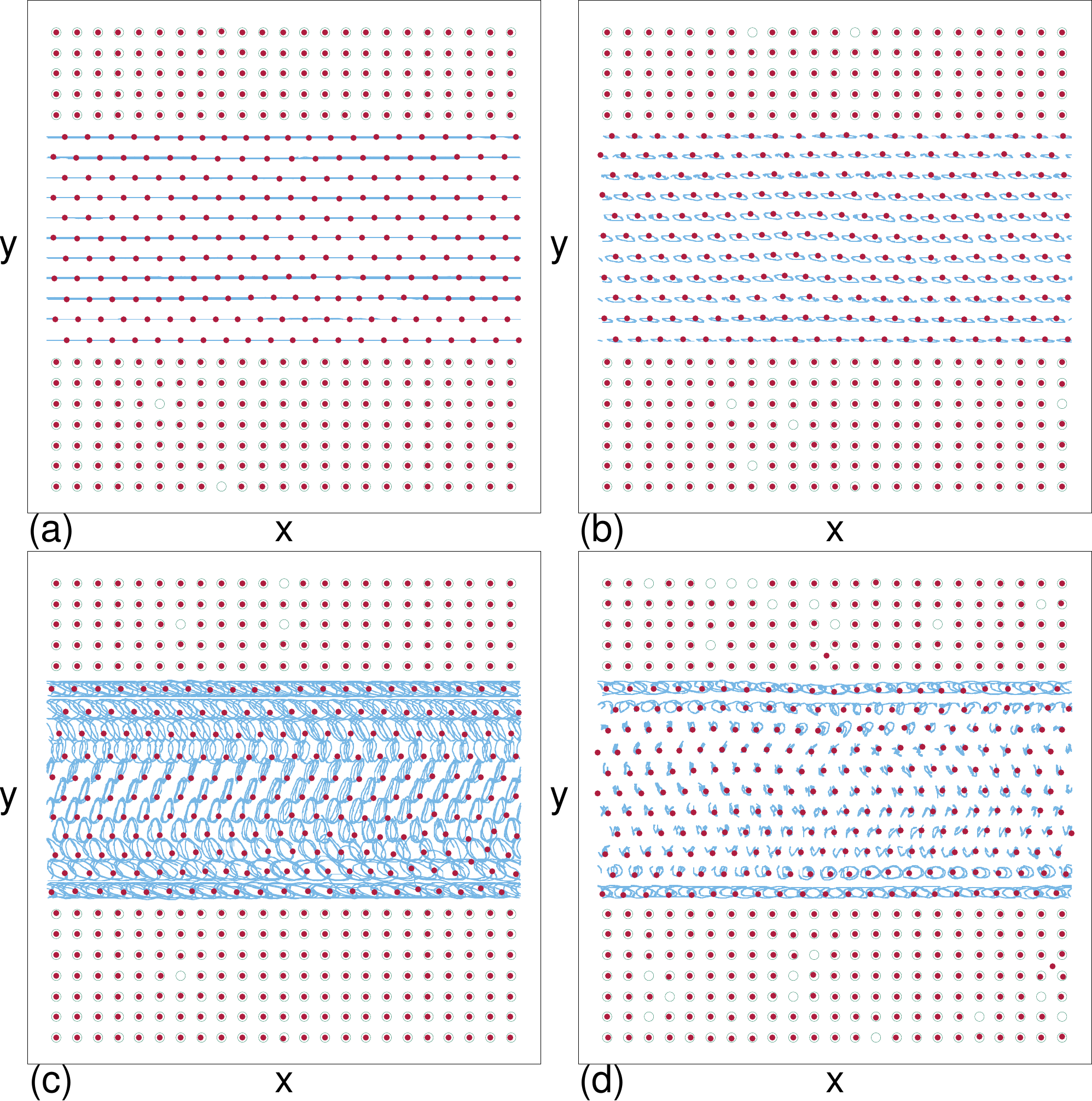}
\caption{Skyrmion positions (dots), trajectories (lines), and
  pinning site locations (open circles)
  for varied ratios of the Magnus force to the damping term
  in samples with $x$ direction driving and $n_{sk}=n_{\phi}=0.4$.
  (a) At $A = 0.075$ and $\omega = 3.75 \times 10^{-5}$ for
  $\alpha_{m}/\alpha_{d} = 0.0$ in the overdamped limit,
  there is no edge transport and the skyrmions move in strictly 1D orbits.
  (b)
  At $A = 0.025$ and $\omega = 1.5 \times 10^{-4}$ for
  $\alpha_{m}/\alpha_{d} = 0.5$,
  there is no edge transport
  and the skyrmions move in elliptical paths.
  (c) At $A = 0.075$ and $\omega = 3.75 \times 10^{-5}$ for
  $\alpha_{m}/\alpha_{d} = 5$,
  there are multiple rows participating in the edge transport.
  (d) At $A = 0.075$ and $\omega = 3.75 \times 10^{-5}$ for
  $\alpha_{m}/\alpha_{d} = 20$,
  the edge transport involves only the edgemost rows
  and the skyrmion orbits are smaller.}
\label{fig:6}
\end{figure}

\begin{figure}
\includegraphics[width=3.5in]{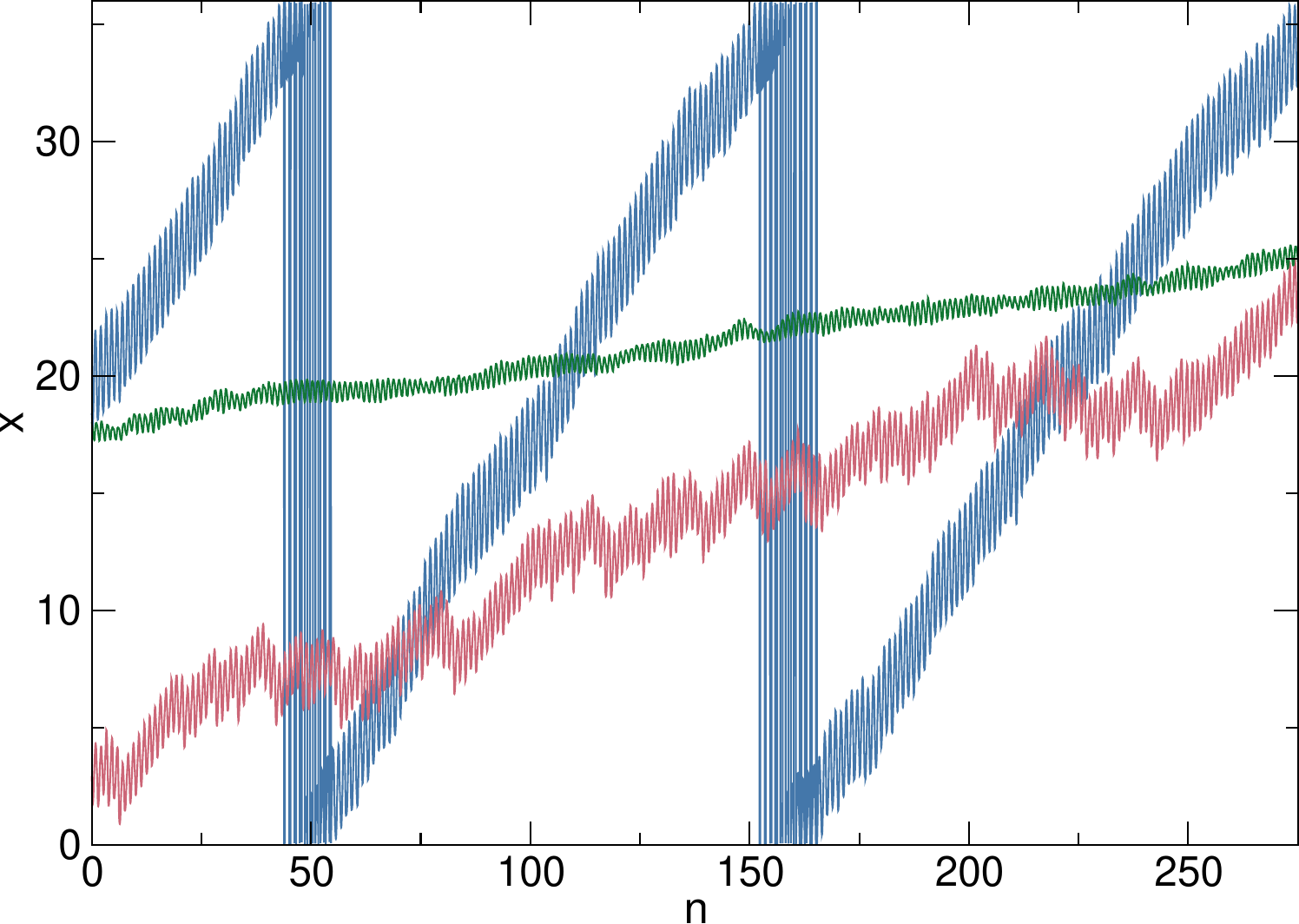}
\caption{$x$ position vs ac cycle number $n$ for
  individual skyrmions in the first (blue), second (red), and third (green) rows from the
  top of the pin-free channel for the sample
  in Fig.~\ref{fig:6}(c) with $x$ direction driving at
  $n_{sk}=n_{\phi}=0.4$,
  $\alpha_{m}/\alpha_{d} = 5.0$,
  $\omega = 3.75 \times 10^{-5}$, and $A = 0.075$. 
  Multiple rows are moving but the transport is reduced for rows that are
  further from the edge of the pin-free channel.}
\label{fig:7}
\end{figure}

In Fig.~\ref{fig:6} we illustrate the skyrmion trajectories for
varied $\alpha_{m}/\alpha_{d}$. 
In the overdamped limit of
$\alpha_{m}/\alpha_{d} = 0.0$,
shown in Fig.~\ref{fig:6}(a)
for $A=0.075$ and $\omega=3.75 \times 10^{-5}$,
there are no edge currents and the skyrmions in the
pin-free channel
move in strictly 1D paths aligned with the 
$x$-direction.
Here the orbits are large enough that they overlap.
In Fig.~\ref{fig:6}(b),
the orbits for
a system with $\alpha_{m}/\alpha_{d} = 0.5$, $A=0.025$, and
$\omega = 1.5 \times 10^{-4}$
are elliptical but there are still no edge currents.
The sample with $\alpha_{m}/\alpha_{d} = 5.0$, $A=0.075$, and
$\omega = 3.75 \times 10^{-5}$ illustrated in Fig.~\ref{fig:6}(c)
has strong edge transport as well as
transport of skyrmions that are up to three rows from the edge
of the pin-free channel.
This
occurs when the edge transport is sufficiently strong that it exerts
a drag effect on the adjacent rows,
which pick up a net motion at a reduced velocity.
In Fig.~\ref{fig:7} we illustrate
the $x$-position of individual skyrmions in the top, second from top, and third
from top rows in the pin-free channel for the system in Fig.~\ref{fig:6}(c).
Each of these rows has a net transport in the positive $x$ direction, but the magnitude
of the transport decreases as the rows become further from the edge of the
channel.
The forth row from the top of the channel has no net drift.
Similarly, at the bottom of the pin-free channel (not shown) the three rows closest
to the channel edge are moving in the $-x$ direction.
We have also found regimes in which only two rows are moving as well as regimes
in which only the edgemost row is moving.
Figure~\ref{fig:6}(d) shows the
skyrmion trajectories for a sample with $\alpha_{m}/\alpha_{d} = 25$,
$\omega=3.75 \times 10^{-5}$,
and $A = 0.075$.
The orbits are smaller than those
that appear for the same ac drive parameters at
$\alpha_{m}/\alpha_{d} = 15$ or $10$ since
the skyrmion orbits become
more curved when the Magnus force is larger.
When the skyrmion orbits are smaller, the
edge currents are reduced in magnitude.

\begin{figure}
\includegraphics[width=3.5in]{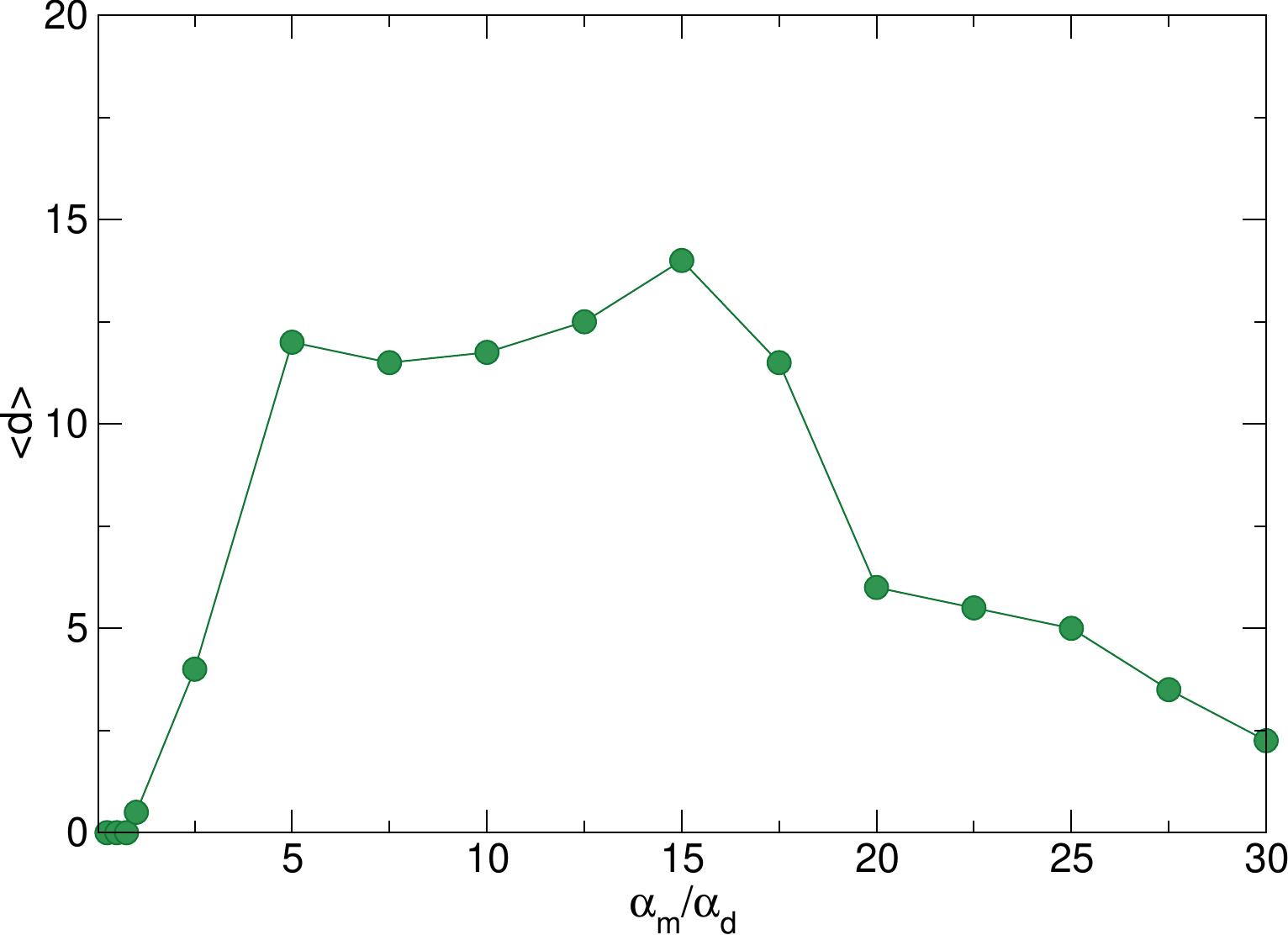}
\caption{ $\langle d\rangle$ vs $\alpha_{m}/\alpha_{d}$ for
  an $x$ direction ac drive with $\omega = 3.75 \times 10^{-5}$
  and $A = 0.075$, showing
  that $\langle d\rangle$ goes to zero
  in the overdamped limit of $\alpha_{m}/\alpha_{d} = 0.0$.
Here $n_{sk}=n_{\phi}=0.4.$}
\label{fig:8}
\end{figure}

In Fig.~\ref{fig:8} we plot $\langle d\rangle$ versus
$\alpha_{m}/\alpha_{d}$ for
an ac drive with $\omega = 3.75 \times 10^{-5}$
and $A = 0.075$.
Edge currents are absent when
$\alpha_{m}/\alpha_{d}$ is small.
For $2.5 < \alpha_{m}/\alpha_{d} < 17.5$,
multiple rows participate in the edge transport,
while for $\alpha_{m}/\alpha_{d} \geq 17.5$,
only the outer row of the pin-free channel has a net transport.
The overall shape of $\langle d\rangle$ versus
$\alpha_{m}/\alpha_{d}$ depends strongly 
on the values of $\omega$ and $A$,
and if these quantities are too small, $\langle d\rangle = 0.0$.

\begin{figure}
\includegraphics[width=3.5in]{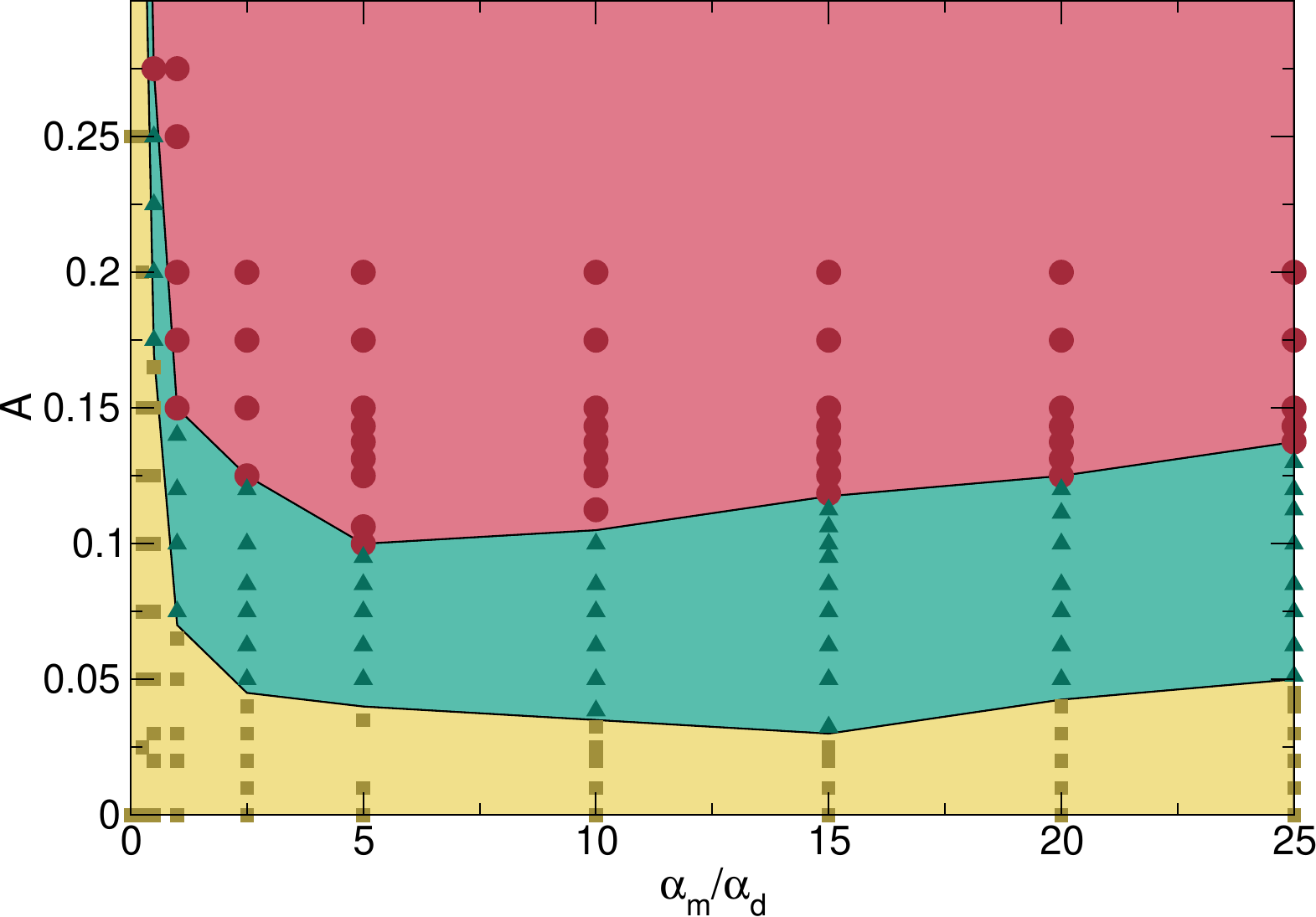}
  \caption{Dynamic phase diagram as a function of $A$ vs $\alpha_{m}/\alpha_{d}$ for
    fixed $\omega = 3.75 \times 10^{-5}$ and $x$ direction driving in
    samples with $n_{sk}=n_{\phi}=0.4$.
    Squares (yellow): localized state;
    triangles (green): crystal state with edge currents;
    circles (red): disordered fluctuating state.
} 
\label{fig:9}
\end{figure}

By varying $A$, $\omega$, and $\alpha_m/\alpha_d$,
we identify three regimes of behavior:
a localized lattice phase
with no edge transport,
an edge transport phase in which the skyrmions remain in a lattice structure,
and a disordered or fluctuating state.
In Fig.~\ref{fig:9} we construct a dynamic phase diagram as a function
of $A$ versus $\alpha_m/\alpha_d$ 
for a system with fixed $\omega = 3.75 \times 10^{-5}$,
highlighting the localized phase,
lattice edge transport phase, and fluctuating liquid phase.
As $\alpha_{m}/\alpha_{d}$ approaches zero,
the localized phase grows in extent. 
In the overdamped limit, the skyrmions form a
triangular lattice that moves elastically back
and forth in the $x$-direction
with no net transport, as illustrated
in Fig.~\ref{fig:6}(a).
Figure~\ref{fig:9} indicates that
there is a tendency for the localized region
to grow at large values of $\alpha_{m}/\alpha_{d}$
due to the shrinking of the skyrmion orbit size with increasing Magnus force.
Within the fluctuating regime,
there is still some transport
of skyrmions along the edges
at smaller values of $A$;
however, for larger $A$ the system becomes more
liquid like
and the edge transport vanishes.
We can construct a similar phase diagram
for fixed $A$ and varied $\omega$ (not shown), where  we find that
at high drive frequencies, the system enters the localized regime.

\section{AC driving in the Perpendicular Direction}

Up until now we have considered a driving force
applied along the $x$-direction, parallel to the pin-free channel.
In this orientation, the Magnus force generated by the drive produces
a skyrmion velocity component aligned mostly along the $y$-direction.
If the ac drive is instead applied along the $y$-direction, perpendicular
to the pin-free channel, the
Magnus-induced velocity
is mostly aligned with the $x$-direction,
and as a result,
for large $\alpha_{m}/\alpha_{d}$,
the skyrmion trajectories
become nearly 1D along the $x$-direction and the edge currents are absent.

\begin{figure}
\includegraphics[width=3.5in]{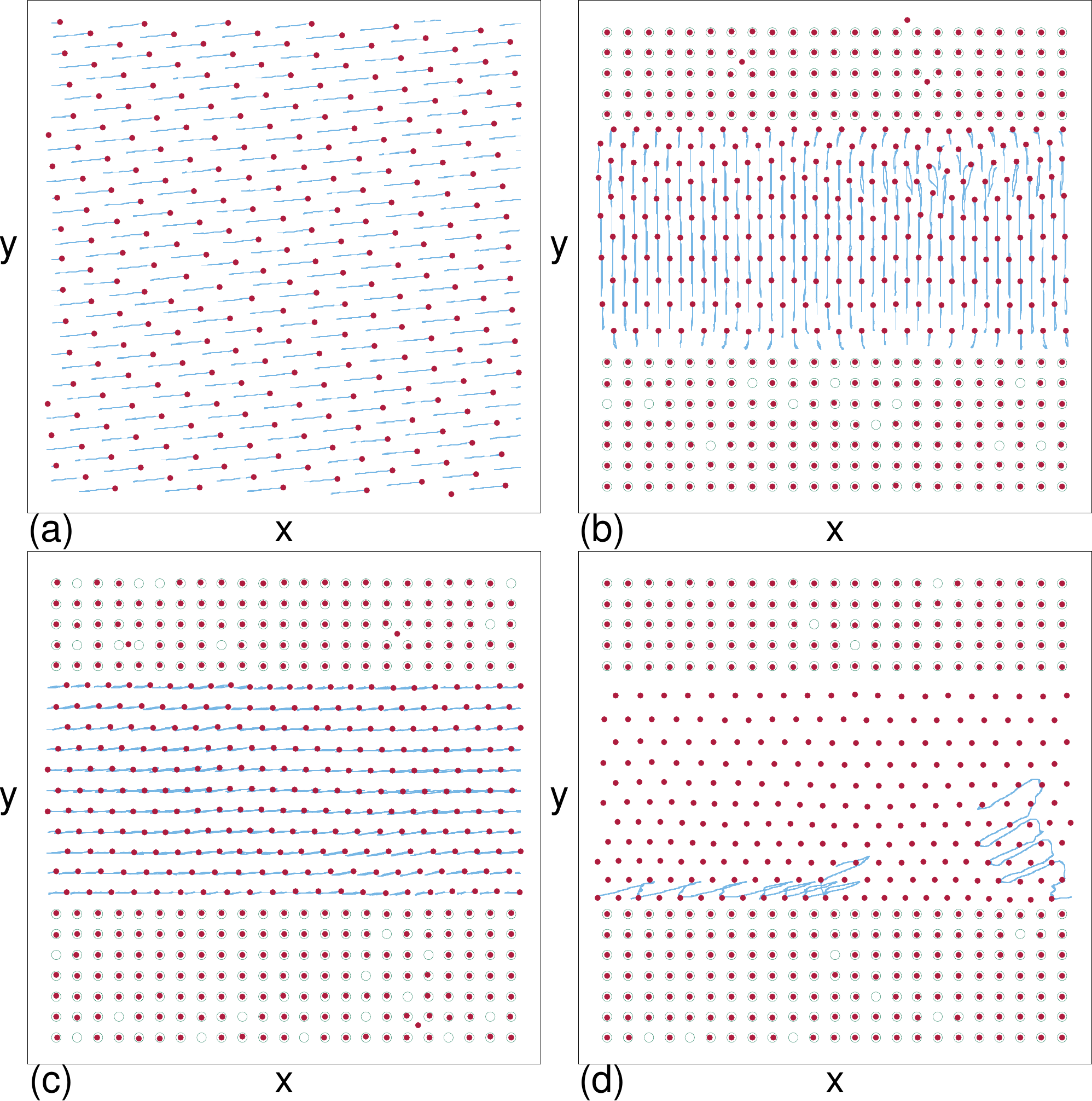}
\caption{ Skyrmion positions (dots), trajectories (lines), and pinning site locations
  (open circles) for systems with $n_{sk}=n_{\phi}=0.4$ in which
  the ac drive is applied in the $y$-direction.
  (a) A sample with no pinning at $A = 0.075$, $\omega = 3.75 \times 10^{-5}$
  and 
  $\alpha_{m}/\alpha_{d} = 10$.
  The trajectories are 1D in nature and are aligned
  mostly along the $x$-direction.
  (b) The same sample with pinning at
  $\alpha_{m}/\alpha_{d} = 0.0$
  for
  $A = 0.075$ and $\omega = 3.75 \times 10^{-5}$,
  where there are no edge currents.
  (c) At $\alpha_{m}/\alpha_{d} = 10.0$
  for
  $A = 0.075$ and $\omega = 3.75 \times 10^{-5}$,
  edge currents appear.
  (d) A sample with $\alpha_{m}/\alpha_{d} = 2.0$, $A = 0.075$,
  and $\omega = 5 \times 10^{-6}$ in the fluctuating state
  where the trajectory of only a single skyrmion is plotted.
  This skyrmion
  undergoes edge transport before becoming trapped in the bulk of
  the pin-free channel.
} 
\label{fig:10}
\end{figure}

In Fig.~\ref{fig:10}(a)
we illustrate the skyrmion trajectories
in the absence of pinning for a system with
an ac drive applied along the $y$-direction
at $A = 0.075$, $\omega = 3.75 \times 10^{-5}$,
and $\alpha_{m}/\alpha_{d} = 10$.
Here the skyrmions follow
1D paths
that are aligned mostly in the $x$-direction.
Figure~\ref{fig:10}(b) shows the same system with pinning
present in the overdamped limit of
$\alpha_{m}/\alpha_{d} = 0.0$.
The skyrmions in the pin-free channel
move in 1D paths aligned in the $y$-direction
and there is no edge transport.
In Fig.~\ref{fig:10}(c) at $\alpha_{m}/\alpha_{d} = 10$, 
the skyrmions move in 
mostly 1D  paths aligned with the $x$-direction,
and edge transport is absent.
In general, for driving in the $y$-direction
we only observe edge transport for low values
of $\alpha_{m}/\alpha_{d}$ and low
ac drive frequencies $\omega$.
The edge transport is the most efficient when
the system is near the transition to the fluctuating state.
In Fig.~\ref{fig:10}(d)
at $A = 0.075$, $\alpha_{m}/\alpha_{d} = 2.0$,
and $\omega = 5 \times 10^{-6}$, just past the transition to
the fluctuating state,
we highlight the trajectory of a single skyrmion
that undergoes edge transport when it is near the edge
of the pin-free channel.  The skyrmion
eventually wanders off into the bulk and becomes trapped.

\begin{figure}
\includegraphics[width=3.5in]{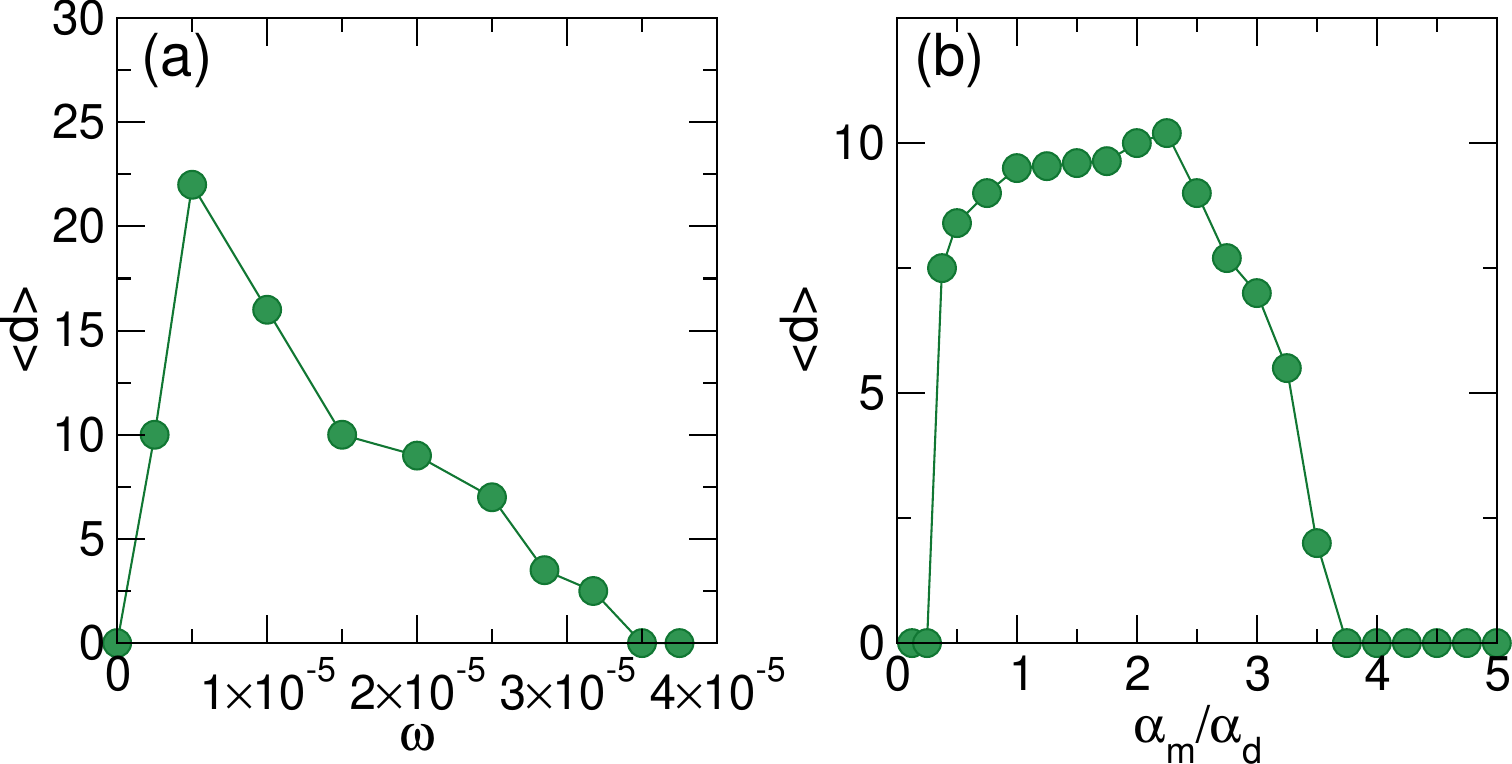}
\caption{
  (a) $\langle d\rangle$ vs $\omega$ at $\alpha_m/\alpha_d=2$
  and (b) $\langle d\rangle$ vs $\alpha_{m}/\alpha_{d}$ at
  $\omega=5 \times 10^{-6}$
  for the system in Fig.~\ref{fig:10}(d)
  with $y$ direction driving, $n_{sk}=n_{\phi}=0.4$, and $A=0.075$.
} 
\label{fig:11}
\end{figure}

In Fig.~\ref{fig:11}(a,b) we plot
$\langle d\rangle$ versus $\omega$ and
$\langle d\rangle$ versus $\alpha_{m}/\alpha_{d}$, respectively,
for the system in Fig.~\ref{fig:10}(d).
The frequency and amplitude dependence of the edge currents is
similar to that found for driving in the $x$-direction.
Here there are no edge currents for 
$\alpha_{m}/\alpha_{d} > 3.5$ since the orbits
become too one-dimensional.

\section{Effects of Disorder}

\begin{figure}
  \includegraphics[width=3.3in]{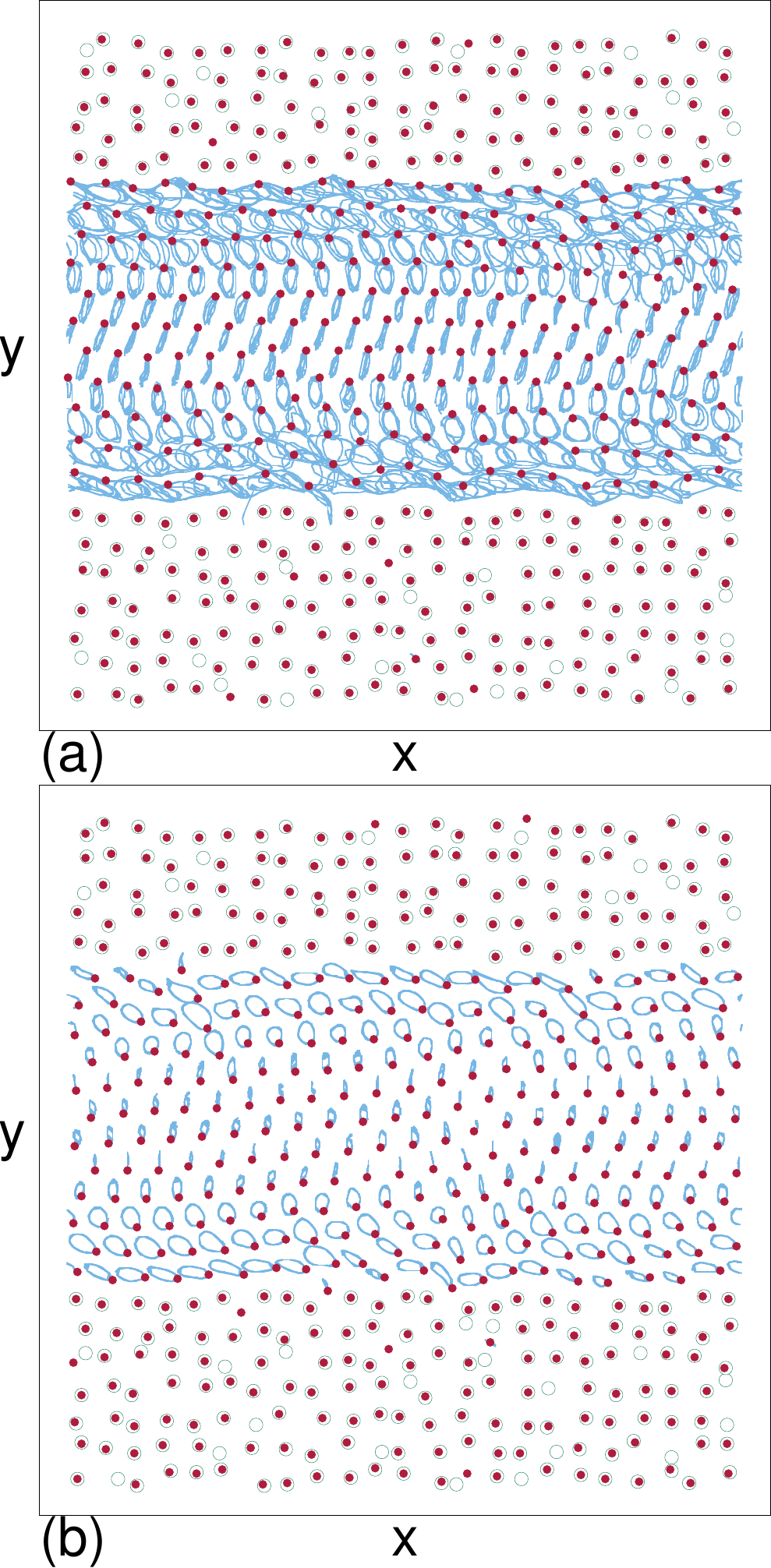}
\caption{Skyrmion positions (dots), trajectories (lines), and pinning site
  locations (open circles) for samples in which disorder has been added in the form
  of random offsets $\delta r$ of the pinning site locations.
  Here, the driving is in the $x$ direction and $n_{sk}=n_{\phi}=0.4$.
  (a) At $\omega = 1.025 \times 10^{-4}$,
  $A = 0.075$, $\alpha_{m}/\alpha_{d} = 15$, and $\delta r = 0.5$,
  there is disorder-induced edge transport.
  (b) In the same system at
  $\omega = 1.5 \times 10^{-4}$,
  there is no edge transport. 
}
\label{fig:12} 
\end{figure}

\begin{figure}
\includegraphics[width=3.5in]{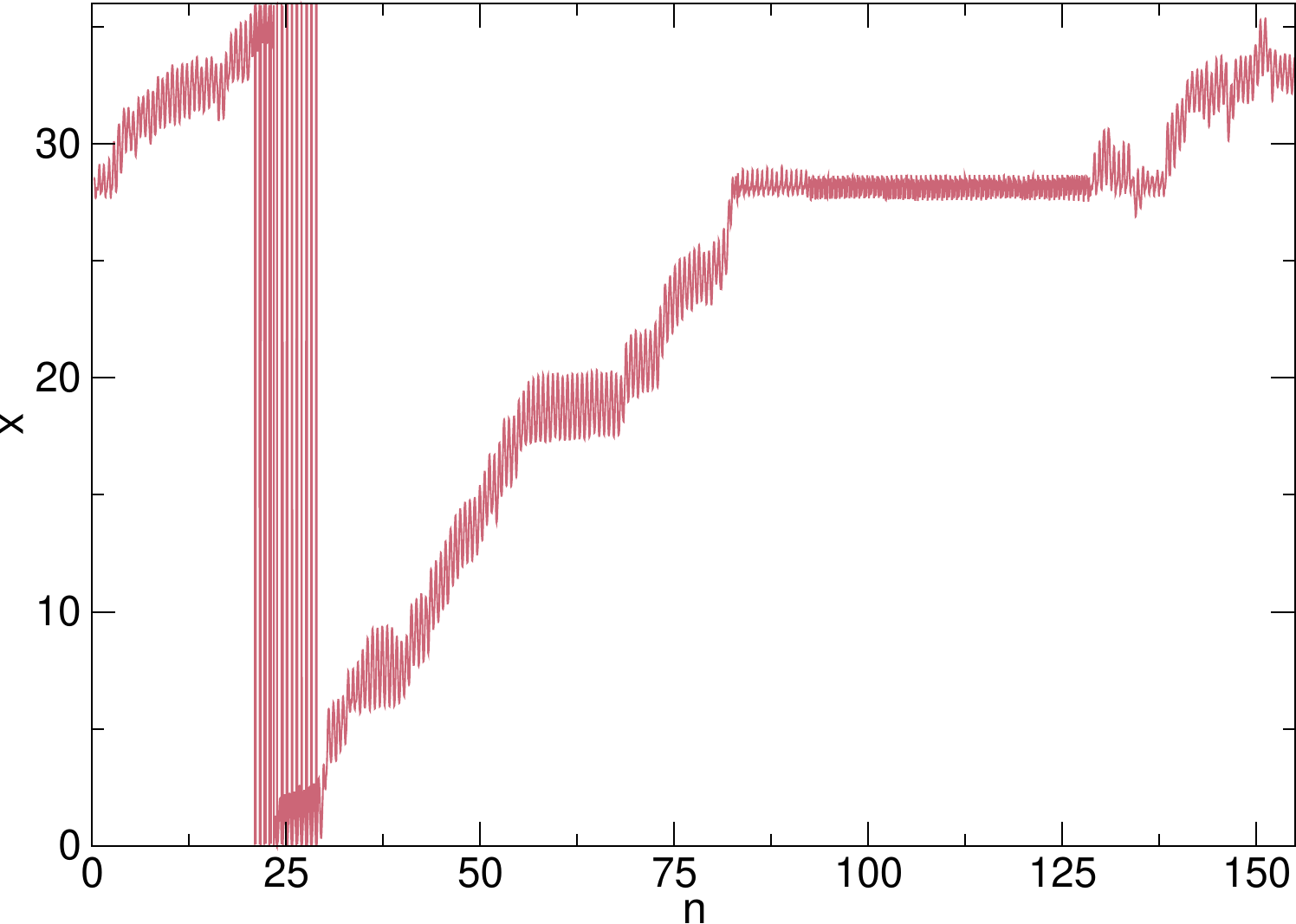}
\caption{ $x$ position vs ac cycle number $n$ for a single skyrmion in a system with
  $x$ direction driving, $n_{sk}=n_{\phi}=0.4$,
  $\omega = 3.75 \times 10^{-5}$,
  $A = 0.075$, $\delta r = 0.5$, and $\alpha_{m}/\alpha_{d} = 10$.
  The disorder
  causes the skyrmion to become localized for extended periods of time.
} \label{fig:13}
\end{figure}

We next consider the addition of disorder,
achieved by applying random 
offsets to the pinning site locations in
both the $x$ and $y$ directions.
The offsets are uniformly distributed and have a maximum size of
$\delta r$, which is always less than $0.5a$, where $a$ is the pinning lattice
constant.
Depending on the parameters,
we find that the disorder can enhance or decrease the
edge transport. 
In Fig.~\ref{fig:12}(a) we illustrate
the skyrmion positions and trajectories for a system with
$\delta r = 0.5$, $A = 0.075$,
$\omega = 1.025 \times 10^{-4}$,
and $\alpha_{m}/\alpha_{d} = 15$.
At this value of $\omega$,
the ordered system has no edge transport,
as shown in Fig.~\ref{fig:5}(b).
In the presence of sufficiently large disorder, however,
edge transport occurs in the two edgemost rows.
Figure~\ref{fig:12}(b) shows the same
sample at
a higher ac drive frequency of
$\omega= 1.5 \times 10^{-4}$,
where the orbits are small enough 
that the edge currents are lost even in the presence of disorder.

In other cases,
the addition of disorder reduces the edge currents and
produces
intermittent or chaotic motion in which
the edge current
drops to zero for a period time
before
becoming finite again.
In Fig.~\ref{fig:13} we show an example of this
behavior in a sample with
$\omega = 3.75 \times 10^{-5}$,
$A = 0.075$, $\delta r = 0.5$,
and $\alpha_{m}/\alpha_{d} = 10$, 
which exhibits
pronounced edge transport when $\delta r = 0.0$.
The plot of the $x$ position of a single skyrmion versus ac cycle number $n$ shows
that there are several time intervals
during which the skyrmion becomes localized.
The edge skyrmions trace the same path during each ac cycle in the
localized interval, but this path is chaotic, allowing the skyrmions 
eventually to
jump back into a translating orbit.

\begin{figure}
\includegraphics[width=3.5in]{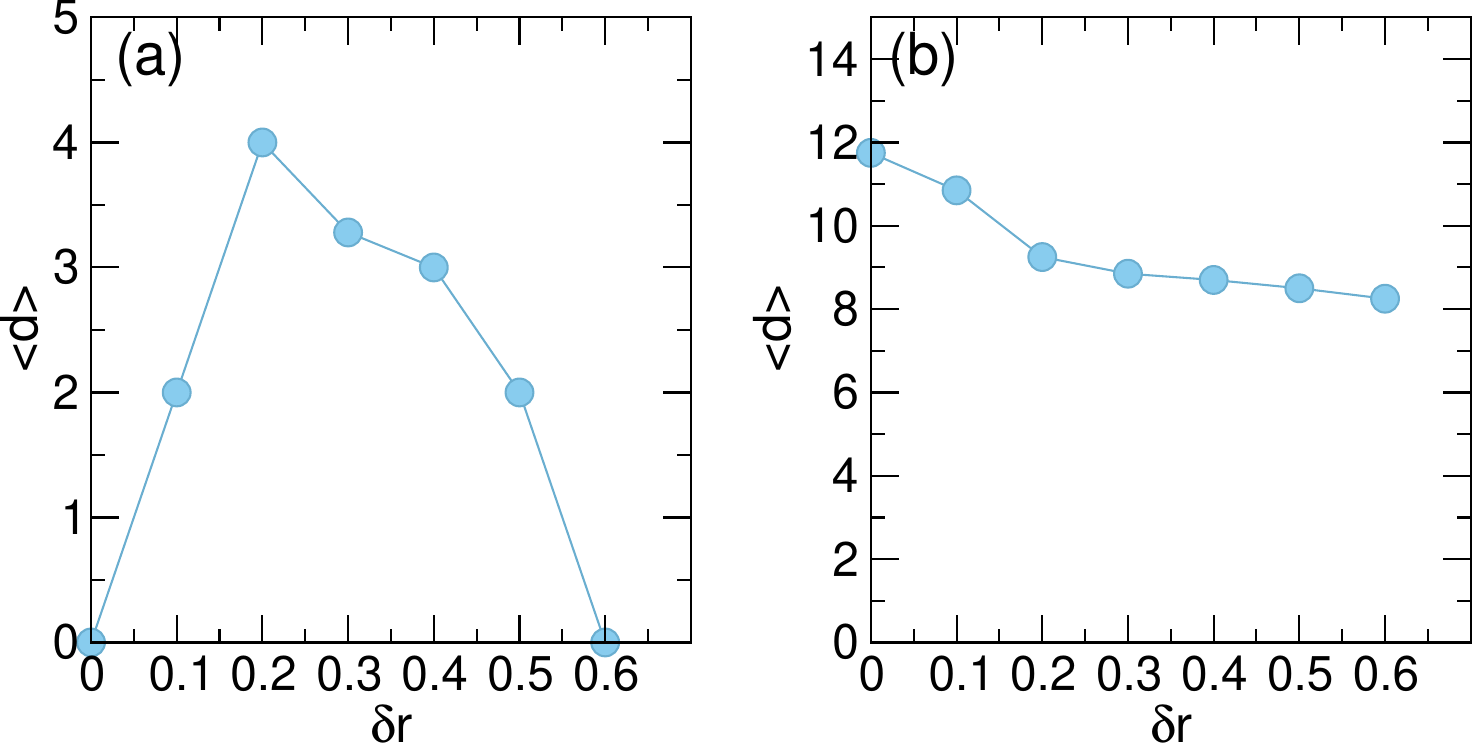}
\caption{  $\langle d\rangle$ vs $\delta r$ for the system
  in Fig.~\ref{fig:12}(a)
  with $x$ direction driving showing that addition of disorder can induce an edge current.
  Here
  $n_{sk}=n_{\phi}=0.4$, $\omega=1.025 \times 10^{-4}$,
  $A=0.075$, and $\alpha_m/\alpha_d=15$.
  (b) $\langle d\rangle$ vs $\delta r$
  for the system in Fig.~\ref{fig:13}
  at $n_{sk}=n_{\phi}=0.4$, $\omega=3.75 \times 10^{-5}$,
  $A=0.075$, and $\alpha_m/\alpha_d=10$.
  where
  addition of disorder reduces the edge current.  
} \label{fig:14}
\end{figure}

In Fig.~\ref{fig:14}(a) we plot $\langle d\rangle$ versus $\delta r$ for the system in
Fig.~\ref{fig:12}(a), where we find an increase in the
edge current
for intermediate values of $\delta r$.
For larger disorder, however,
the transport is reduced 
due to the pinning of the edge current by the disorder.
In Fig.~\ref{fig:14}(b), we show $\langle d\rangle$ versus $\delta r$
for the system in Fig.~\ref{fig:13} where
the addition of disorder decreases the
edge transport. We find similar effects for driving 
in  the $y$-direction.  
  
\section{Density and Skyrmion Pumping Effects}

\begin{figure}
\includegraphics[width=3.5in]{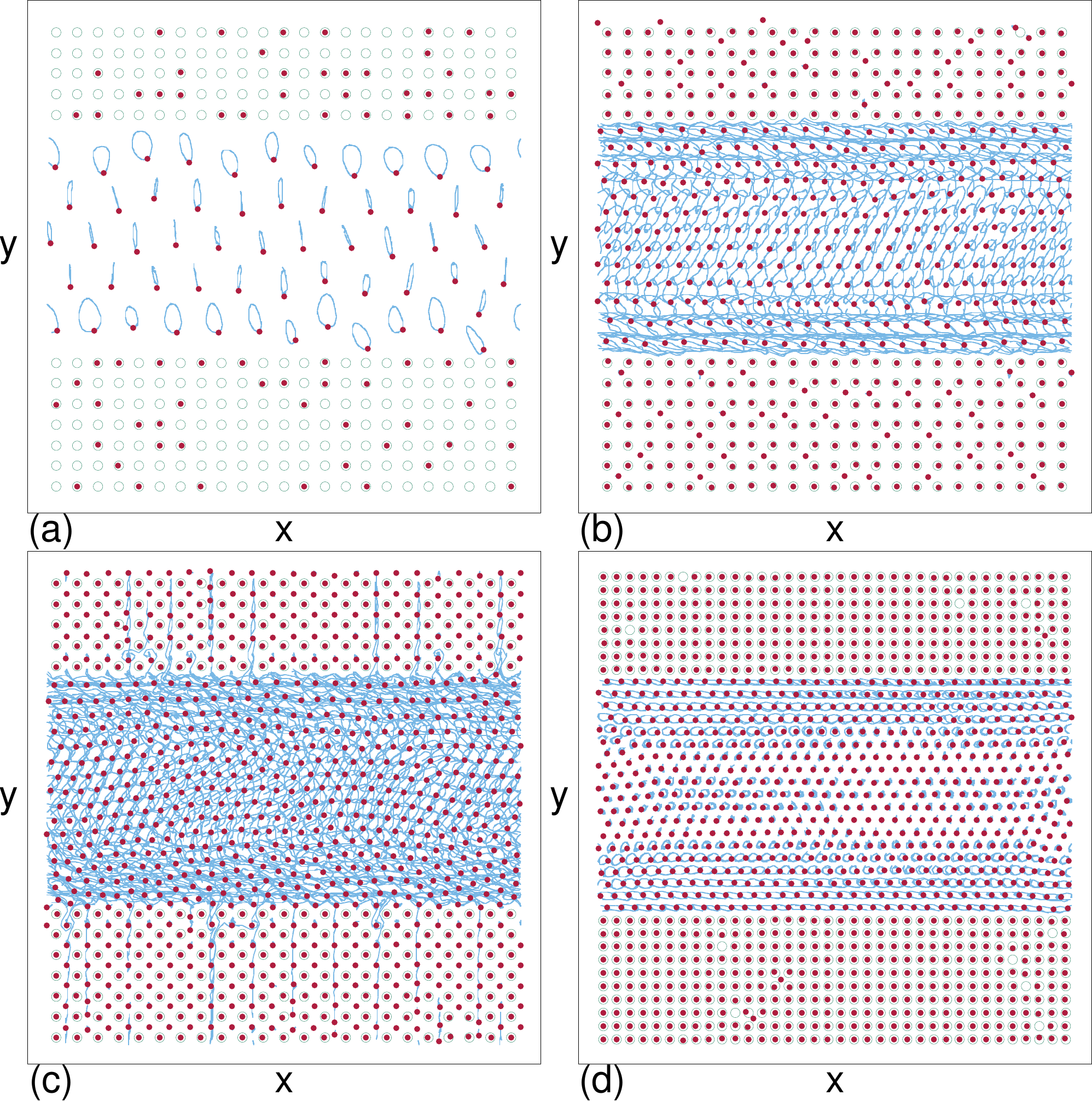}
\caption{Skyrmion positions (dots), trajectories (lines), and pinning site
  locations (open circles)
  for a system with
  $x$ direction driving at $A = 0.075$, $\omega = 3.75 \times 10^{-5}$,
  and $\alpha_{m}/\alpha_{d} = 10$ for fixed matching density
  $n_{\phi}=0.4$ and varied
  skyrmion density.
  (a) At $n_{sk} = 0.1$, there is no edge current.
  (b) At $n_{sk} = 0.5$,  there is strong edge transport.
  (c) At $n_{sk} = 1.0$, there is considerable
disorder in the trajectories and the edge transport is reduced.  
(d) A system with a
matching density of $n_{\phi}=1.0$ and skyrmion density
$n_{sk} = 1.0$ for $A = 0.25$, $\omega=3.75 \times 10^{-5}$,
and $\alpha_{m}/\alpha_{d} = 10$, showing edge transport.  
}
\label{fig:15}
\end{figure}

\begin{figure}
\includegraphics[width=3.5in]{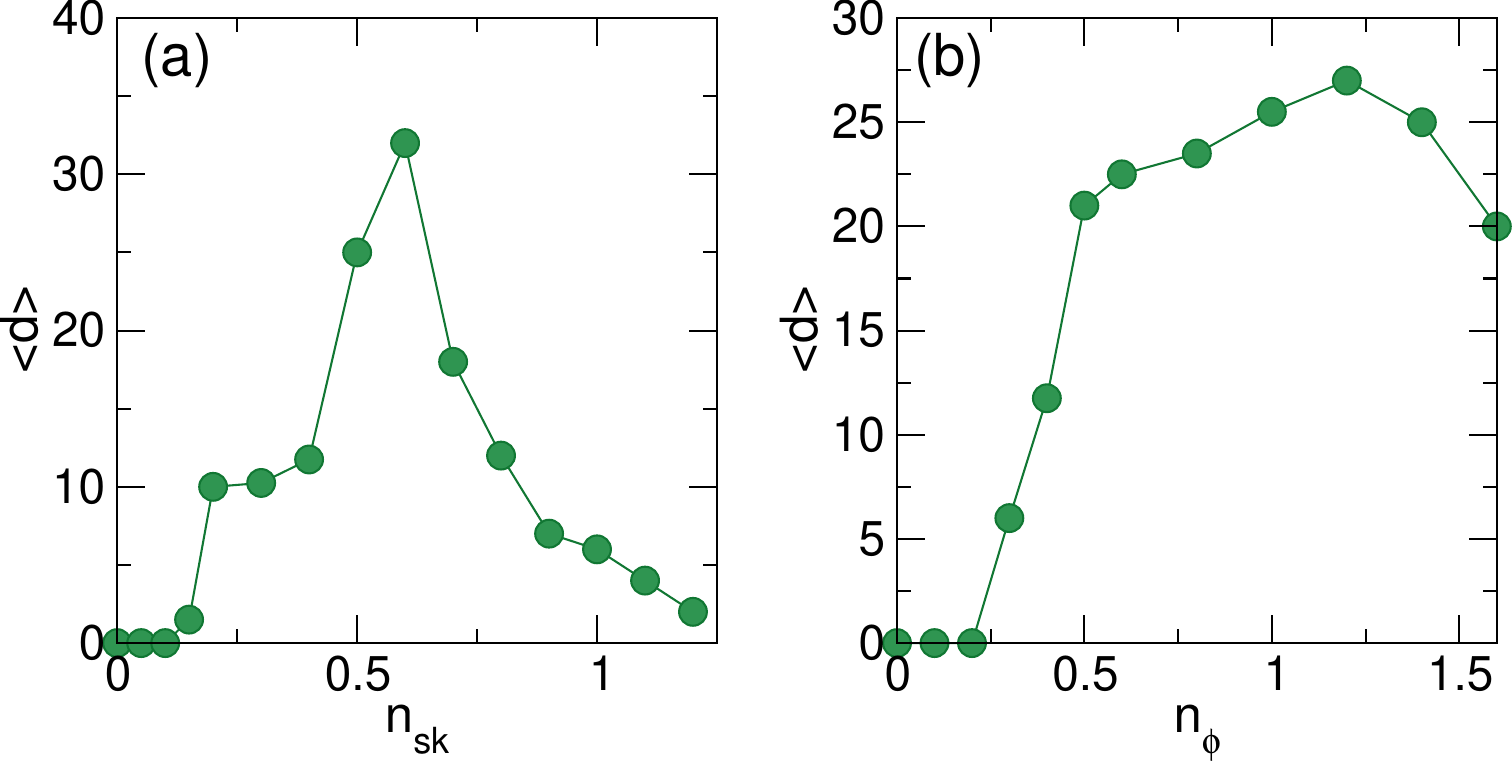}
\caption{ (a) $\langle d\rangle$ vs $n_{sk}$ for a system with a fixed
  matching density of $n_{\phi} = 0.4$ under $x$ direction driving
  at $\omega = 3.75 \times 10^{-5}$,
  $A = 0.075$, and $\alpha_{m}/\alpha_{d} = 10$.
  The edge currents are lost at low skyrmion density and diminish with
  increasing $n_{sk}$ at high skyrmion density.
  (b) $\langle d\rangle$ vs the matching density $n_{\phi}$
  in samples with $n_{sk}=n_{\phi}$,
  $\omega=3.75 \times 10^{-5}$, $A=0.075$, and $\alpha_m/\alpha_d=10$,
  showing that the edge
  transport is robust over a wide range of system densities. }
\label{fig:16}
\end{figure}

To determine the effects of changing
the skyrmion density,  we study both the case in which the pinning density is held
fixed and the skyrmion density is varied, as well as the case in which the ratio of the
number of pins to the number of skyrmions is held fixed at
$N_p/N_{sk}=1/2$ but the matching density $n_{\phi}$
of the system is changed.
In Fig.~\ref{fig:15}(a) we plot the skyrmion trajectories at
$\alpha_{m}/\alpha_{d} = 10$, $A = 0.075$, and $\omega = 3.75 \times 10^{-5}$
at a skyrmion density of $n_{sk} = 0.1$
and a fixed matching density of $n_{\phi}=0.4$. 
The skyrmion density is low enough that no edge transport occurs. 
Figure~\ref{fig:15}(b) shows the same
system at $n_{sk} = 0.5$ where there is strong edge transport, 
and Fig.~\ref{fig:15}(c) illustrates
the trajectories
at $n_{sk} = 1.0$, where the edge transport is reduced and there is an increase in
the motion of
interstitial skyrmions within the pinned region. 
In Fig.~\ref{fig:15}(d),
the matching density is increased to $n_{\phi}=1.0$
and the skyrmion density is $n_{sk}=1.0$.
Here, at $A = 0.025$ and $\omega=3.75 \times 10^{-5}$,
edge currents are present.  
In Fig.~\ref{fig:16}(a) we plot $\langle d\rangle$ versus $n_{sk}$
for the system in Fig.~\ref{fig:15} with a fixed matching density 
of $n_{\phi} = 0.4$.
For $n_{sk} < 0.15$, there
is no edge current,
while for $0.4 < n_{sk} < 0.8$, pronounced edge transport appears
which falls off when $n_{sk} > 0.8$.
The decrease in the edge transport with increasing $n_{sk}$ at higher
densities
occurs because the number of interstitial skyrmions in the pinned region
is increasing,
contributing a fluctuating component of increasing magnitude to the
edge potential.
In addition, at higher skyrmion densities,
the skyrmion orbits become compressed due to the decrease in
the skyrmion lattice constant. 
This effect is similar to what we find
for reduced ac amplitude or higher ac drive frequencies, both of which
suppress the edge transport.
The peak value in $\langle d\rangle$ is produced by
a resonance effect
in which
the width of the skyrmion orbits along the channel edge
locks
to the periodicity of the
confining potential created by the pinned skyrmions.

\begin{figure}
\includegraphics[width=3.5in]{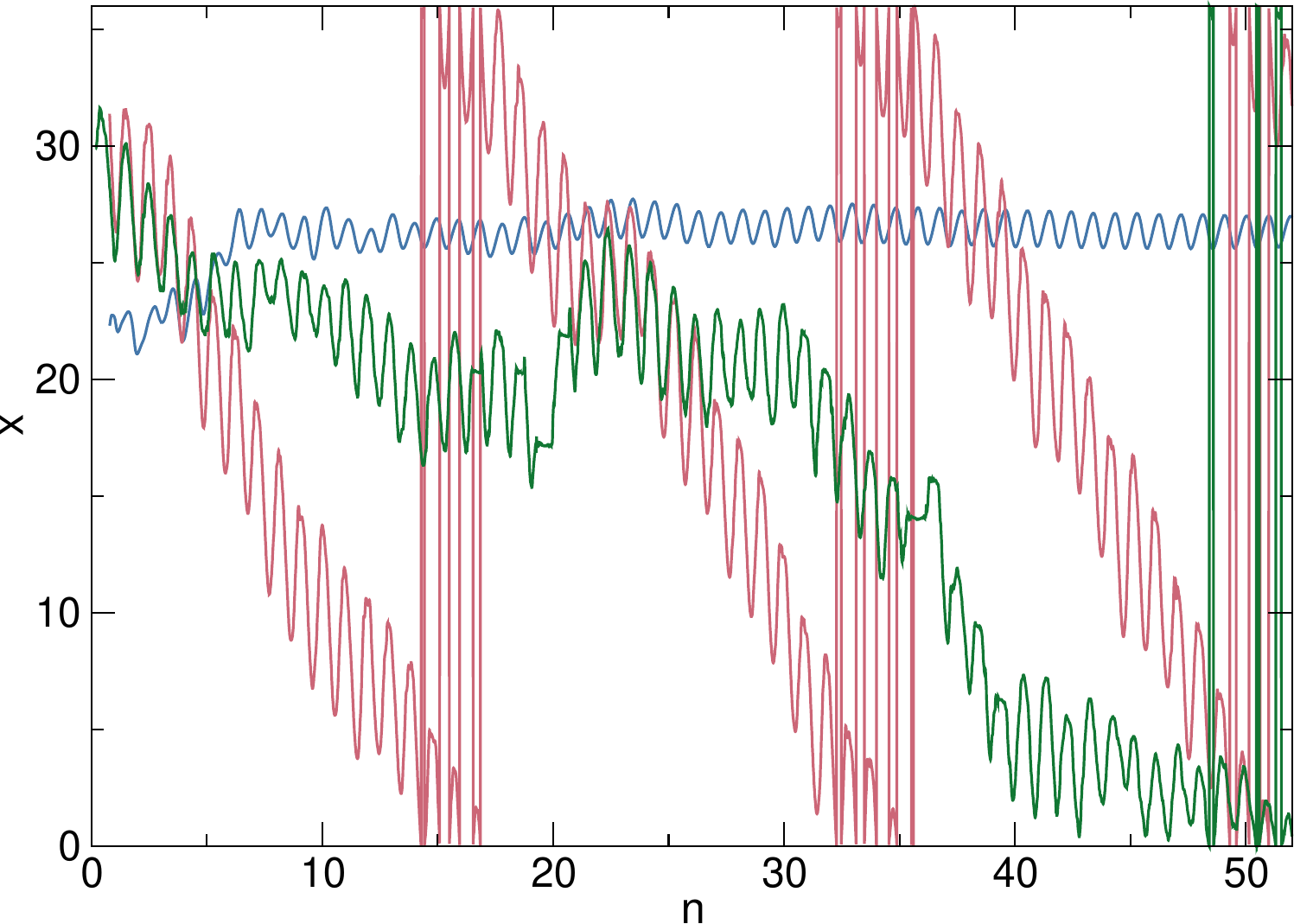}
\caption{ $x$ positions vs ac cycle number $n$ for an individual
  skyrmion on the lower edge of the pin-free channel
  in the system from Fig.~\ref{fig:15}(a,b,c)
  with $x$ direction driving at
  $A=0.075$, $\omega=3.75 \times 10^{-5}$, $\alpha_m/\alpha_d=10$,
  and $n_{\phi}=0.4$.
  When $n_{sk} = 0.1$ (blue), the skyrmion is localized.
  For $n_{sk} = 0.5$ (red) there is strong edge transport,
  while at $n_{sk} = 1.0$ (green) the edge transport is reduced. }
\label{fig:17}
\end{figure}

In Fig.~\ref{fig:17} we plot the
$x$ position of a single skyrmion on the lower edge of the pin-free region in
the system
from Fig.~\ref{fig:15}(a,b,c) with $n_{\phi}=0.4$ at three different skyrmion densities.
At $n_{sk} = 0.1$, the skyrmion undergoes a brief transient motion before becoming
localized.
When $n_{sk} = 0.5$,
the skyrmion translates rapidly,
moving nearly one pinning lattice constant per ac cycle,
giving a transport that is close to optimal.
At $n_{sk} = 1.0$, the edge current is strongly reduced.

In Fig.~\ref{fig:16}(b)
we plot $\langle d\rangle$ versus the matching density $n_{\phi}$ in samples
with $n_{sk}=n_{\phi}$ for the same parameters as in Fig.~\ref{fig:16}(a).
Here, a finite edge current
appears only when $n_{\phi} > 0.3$, and the edge transport remains robust
as $n_{\phi}$ is further increased.
This result indicates
that edge currents should be
a general effect which can be observed whenever
collective interactions between the skyrmions are important.

\begin{figure}
\includegraphics[width=3.5in]{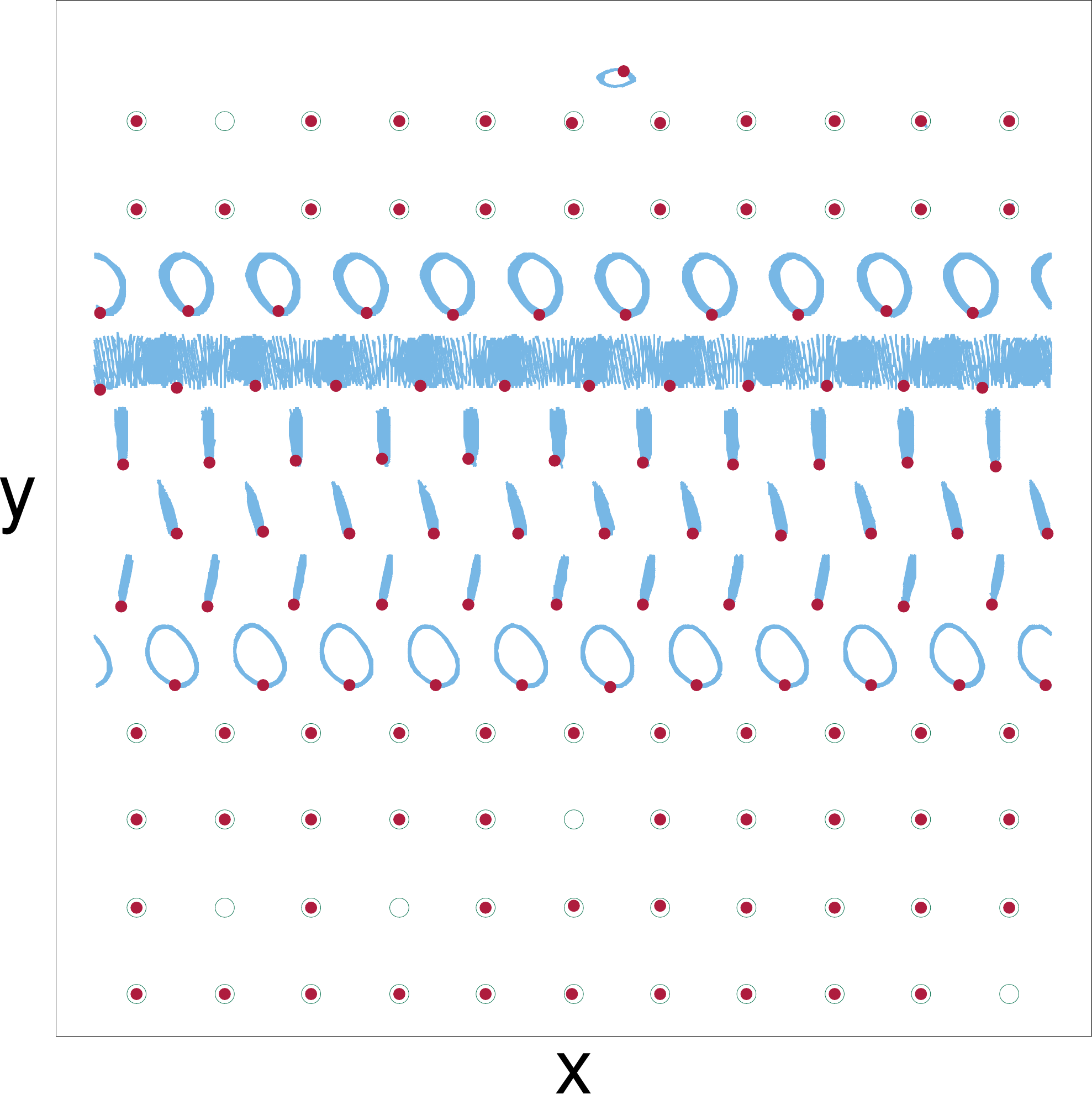}
\caption{Skyrmion positions (dots), trajectories (lines), and
  pinning site locations (open circles) for a system with
  $x$ direction driving and
  a matching density of $n_\phi = 0.1$ at $n_{sk}=0.1$,
  $A = 0.075$, $\omega = 3.75 \times 10^{-5}$,
  and $\alpha_{m}/\alpha_{d} = 10$.
  A different type of current, termed a skyrmion pump effect, occurs
  in the second row from the top of the pin-free channel.
  The skyrmion pump effect appears
  due to an incommensuration between this second row and the
  row at the upper edge of the pin-free channel.
  The skyrmions are propagating in the negative $x$-direction,
  opposite to the edge current motion that appears for
  higher $n_{\phi}$.
} \label{fig:18}
\end{figure}

\begin{figure}
\includegraphics[width=3.5in]{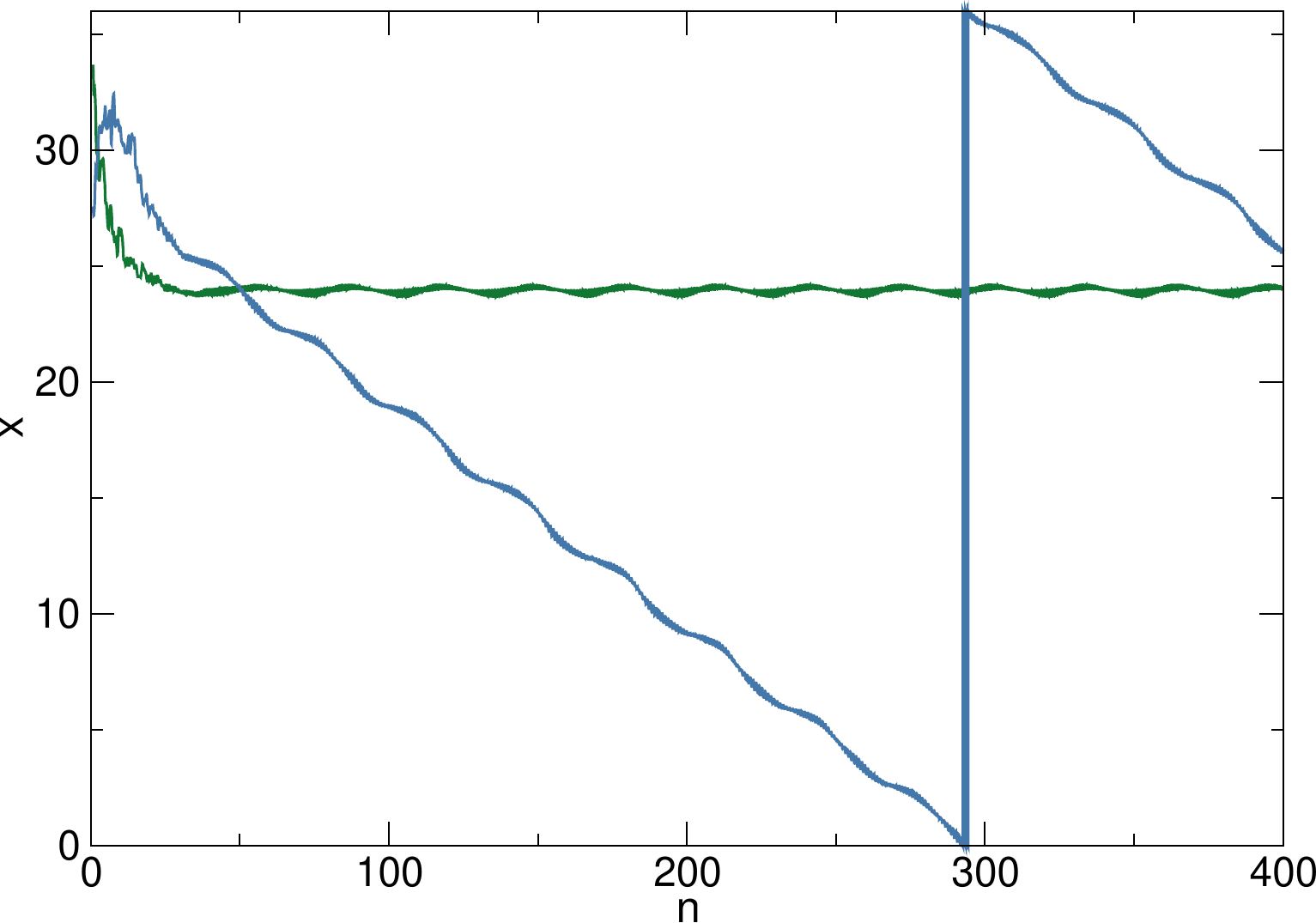}
\caption{$x$ position vs ac cycle number $n$ for
  a skyrmion in the second row from the top of the pin-free channel (blue) and
  a skyrmion in the top row (green)
  for the system in Fig.~\ref{fig:18} with $x$ direction driving
  at $n_{sk}=n_{\phi}=0.1$, $A=0.075$, $\omega=3.75 \times 10^{-5}$,
  and $\alpha_m/\alpha_d=10$.
  The pumped skyrmion is moving in the $-x$ direction,
  which is opposite to the direction of motion found for edge transport of the
  top row at higher $n_{\phi}$.}
\label{fig:19}
\end{figure}

Although we do not find any edge transport when
$n_{\phi}<0.3$,
we
observe a different
type
of translating orbit
that we call a skyrmion pump effect.
In Fig.~\ref{fig:18}, we illustrate the pump effect in a sample with 
$n_\phi = 0.1$, $n_{sk}=0.1$,
$A = 0.075$, $\omega = 3.75 \times 10^{-5}$,
and $\alpha_{m}/\alpha_{d} = 10.0$. 
The skyrmion trajectories, obtained over a time period of
$n=250$ cycles,
indicate that the skyrmions in the second row from the top of the pin-free channel
are undergoing transport, while the top row remains localized.
The pump effect flow is much slower than the edge transport motion
found at higher $n_{\phi}$
and it is also oriented in the opposite direction,
with skyrmions moving in the negative $x$ direction instead of in the
positive $x$ direction.
In Fig.~\ref{fig:19} we plot the $x$ position versus ac drive cycle number $n$
for a skyrmion in the second row from the top of the pin-free channel
in the system from Fig.~\ref{fig:18}.
The skyrmion is moving in the negative $x$-direction
and
exhibits $x$ position oscillations which are much smaller than those
found for edge transport.
For comparison, we also plot the $x$ position of a skyrmion in the top row
of the pin-free channel, which
becomes localized after an initial transient motion.
The skyrmion pump effect arises when
the number of skyrmions in the top row of the pin-free channel is
incommensurate with the number of skyrmions in the adjacent row.
Here, the top row contains 11 skyrmions while the second row from the top
contains 12 skyrmions.
In contrast, both of the bottom two rows in the pin-free channel contain 11 skyrmions.
The incommensuration causes the skyrmions in the top row to act as an effective
gear that gradually translates the skyrmions in the second row from the top.
If the number of skyrmions in the adjacent row is smaller rather than larger than the
number of skyrmions in the top row (10 skyrmions instead of 12),
the pump flow direction is reversed.
We observe the pump effect
when $0.05 < n_{\phi} < 0.2$ in samples with $n_{sk}=n_{\phi}$, and it is
always associated with an
incommensuration between the edge
row and an adjacent row.
For small densities $n_{\phi} \leq 0.05$, the skyrmions are so far apart that their
interactions become unimportant and the pump effect disappears.
We note that a similar pump effect
could also occur at higher densities $n_{\phi} \geq 0.2$ where
edge transport appears;
however, since the pump effect is very weak, it is difficult to detect in
the presence of the much stronger edge transport.

\section{Discussion}
Edge states have previously been proposed to occur
in skyrmion systems,
such as for frustrated magnets \cite{Leonov17}
where a dc current can induce motion on the edges of the sample;
however, this effect is very different from the edge transport
we propose here and it occurs due to a different mechanism.
Chiral magnonic edge states for antiferromagnetic skyrmion
crystals have also been proposed \cite{Diaz19}, but these
differ from the edge current and skyrmion
transport that we consider here.
We note that our results do show similarities with
recent studies of
edge transport in chiral active matter or
active spinning systems,
where directed transport can occur near the edge of the sample or
between regions of spinners of opposite chirality
\cite{vanZuiden16}.
There are still several differences in that the chiral active spinning particles
always undergo circular motion,
whereas in the absence of confinement,
skyrmions subjected to an ac drive
move
along 1D paths,
with circular orbits appearing
only
due to the presence of a confining potential or similar quenched disorder.
The pinning potential we consider
is produced by pinned skyrmions;
however, it is possible to create other types of
confining potentials,
such as by using nanoscale pinning sites that repel skyrmions,
by
modifying the materials properties along the edges
or in stripe patterns \cite{Stosic17,Fernandes18},
or by using nanowires with rough edges \cite{Du15a}.

\begin{figure}
  \includegraphics[width=3.3in]{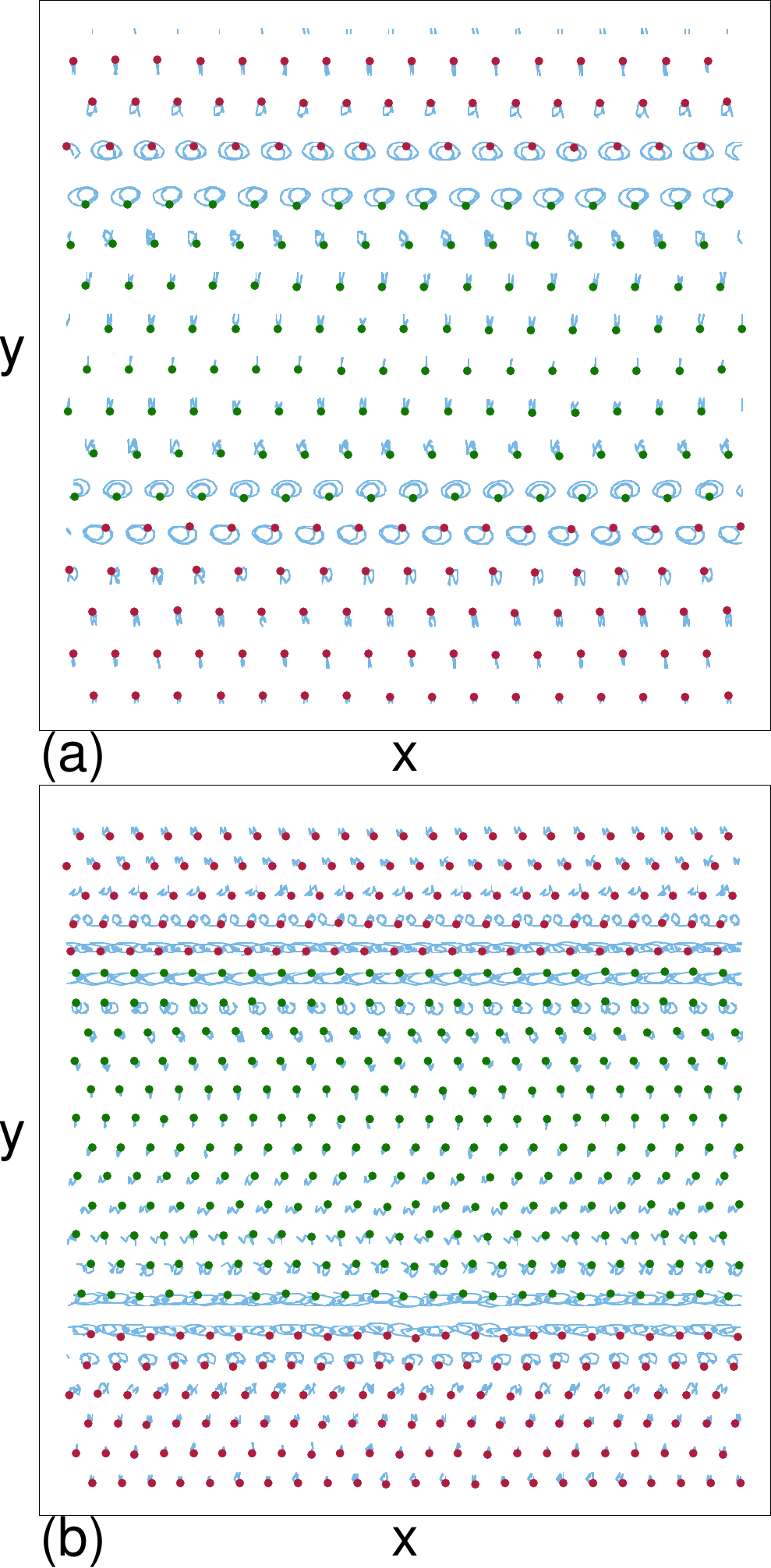}
\caption{Skyrmion positions (dots) and trajectories (lines)
  in a system with
  $x$ direction driving and
  no pinning containing two different skyrmion species
  (red and green) that have Magnus terms of the opposite sign.
  (a) At $n_{sk} = 0.2$, $\alpha_{m}/\alpha_{d} = 10$,
  $A = 0.075$, and $\omega = 1.25 \times 10^{-4}$,
  the largest circular orbits appear
  along the boundary between the two species.
  (b)
  For the same parameters but at $n_{sk} = 0.4$,
  edge currents are flowing along the domain walls separating the two
  species.} \label{fig:20} 
\end{figure}

Beyond nanostructured geometries,
our work suggests that skyrmion edge transport could also be
observed at the interface between
lattices of two different skyrmion species,
between skyrmions with different chiralities \cite{Shibata13},
or along grain boundaries or regions of coexisting
skyrmion and helical states
\cite{Matsumoto16,Li17,Pollath17,Zhang18}. 
In Fig.~\ref{fig:20}(a) we illustrate an example of an edge state that appears
in a pin-free system
at the interface between 
different skyrmion species.
We consider a system with $\alpha_{m}/\alpha_{d} = 10$, $\omega  = 1.25 \times 10^{-4}$,
$A = 0.075$, and $n_{sk}= 0.2$
where half of the skyrmions have a Magnus term that is opposite in sign to the other
half of the skyrmions.  The skyrmions of opposite sign are placed in a band aligned with
the $x$ direction.
Under an ac drive applied in the $x$-direction,
elliptical orbits appear along the domain boundaries separating the two species,
while in the bulk of each domain,
the skyrmions follow 1D trajectories. 
In Fig.~\ref{fig:20}(b) we show the same system at
$n_{sk} = 0.4$, where
clear edge currents emerge at the boundaries between the species.  

Other methods of introducing confining boundaries in the sample include
using inhomogeneous pinning strength so that skyrmions in one region of the sample are
more strongly (or less strongly) pinned than skyrmions in adjacent regions.
It would be
interesting to examine the effects
of confinement on different types of skyrmions,
such as antiferromagnetic skyrmions \cite{Barker16,Legrand19,Nayak17}. 
If the Magnus
force is absent,
the skyrmion dynamics
would be consistent with what is found in
an overdamped system and the edge transport would be absent. 
There are also other
possible systems that could exhibit edge currents,
such as the trochoidal motion of skyrmions in certain bilayer systems \cite{Ritzmann18} 
as well as the dynamics of
anti-skyrmion lattices \cite{Nayak17},
meron lattices  \cite{Yu18}, or polar skyrmions \cite{Das19}.

\begin{figure}
\includegraphics[width=3.3in]{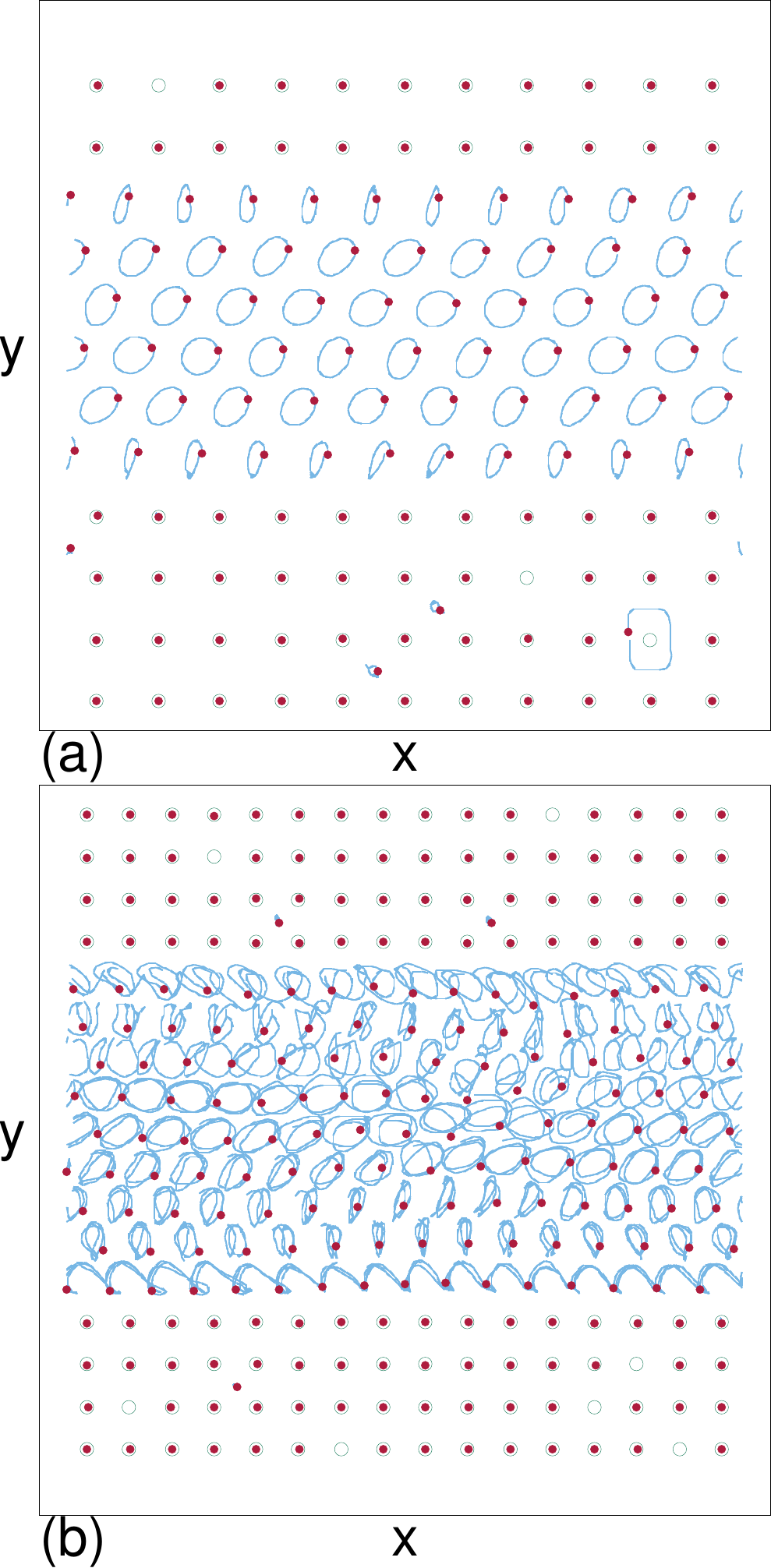}
\caption{ Skyrmion positions (dots), trajectories (lines), and pinning site
  locations (open circles) for a system with a circular ac drive
  applied along the $x$ direction. 
  (a) At $\alpha_{m}/\alpha_{d} = 10$, $A = 0.075$, $\omega = 3.75 \times 10^{-5}$,
  $n_{\phi} = 0.1$, and $n_{sk}=n_{\phi}$,
  the orbits are localized.
  (b) In the same system at $n_{\phi} = 0.2$ and $n_{sk}=n_{\phi}$,
  edge currents are present.}
  \label{fig:21}
\end{figure}

In this work we
considered only a single ac drive applied either parallel or perpendicular to
the pin-free channel.
It is possible, however, to
apply more complex ac drives
such as
${\bf F}^{ac} = A\sin(\omega t){\bf \hat x} + A\cos(\omega t){\bf \hat y}$.
A circular ac drive of this type applied to an underdamped system
can create commentate-incommensurate transitions and
localized or delocalized states in the presence
of periodic or random disorder \cite{Reichhardt02b,Reichhardt06a}.
Addition of an asymmetry in either the substrate or the circular orbit itself
can produce
directed motion in
the overdamped system \cite{Reichhardt03,Soba08,Speer09,Chacon10},
so a rich array of
phenomena is expected to appear for skyrmions under multiple ac drives.
For example, ratcheting skyrmion motion was observed in work on
biharmonic drives
\cite{Chen19}.
In Fig.~\ref{fig:21}(a) we show the trajectories
for a skyrmion system with $n_{\phi} = 0.1$,
$n_{sk}=n_{\phi}$, $\alpha_{m}/\alpha_{d} = 10$, 
$A = 0.075$, and $\omega = 3.75 \times 10^{-5}$
under a circular ac drive.
In this case there
is no edge current;
however, the skyrmions in the bulk of the pin-free channel follow orbits that are
much more circular than the orbits of the skyrmions at the edges of the channel,
in contrast to what we find for a linear ac drive.
In Fig.~\ref{fig:21}(b), the same system at a matching density of
$n = 0.2$ exhibits edge currents, with particularly rapid transport occurring along the
lower edge of the pin-free channel.

Studies of gyroscopic metamaterials
\cite{Nash15,Mitchell18,Mitchell18a}
showed that perturbations
migrate 
to the edge of the sample and propagate around the edge in one direction.
In our system,
for very weak damping or small $\alpha_{m}$,
it would be possible to perturb skyrmions in the bulk
of the pin-free channel
and obtain motion
that is localized on the edges of the channel,
producing a transient current which eventually damps away.
Since the skyrmion lattice
has many similarities to gyroscopic metamaterials,
other effects observed in the latter system 
could be relevant for
skyrmions in confinement,
such as the
odd viscosity and odd elasticity
found
in driven chiral matter \cite{Banerjee17,Soni18,Scheibner19}. 

It would also be possible to use an edge current to create a device.
For example, in a race track geometry filled 
with a skyrmion lattice,
an ac drive could cause a signal to
propagate along the edge of the race track.
One advantage of this mode of operation is
that the propagating skyrmions will
exhibit a skyrmion Hall angle equal to zero, eliminating the problem of having
skyrmions escape from the edges of the track as they move.
Our results
suggest that the propagation speed
of the signal will not monotonically increase with
increasing ac frequency, but will instead
drop to zero at high frequencies.
In our work we have considered only rigid skyrmions,
but in continuum systems,
additional edge modes could arise
such as the prorogation of breathing modes or perturbations of the
internal modes of the skyrmions. 

\section{Summary}

In summary, we have shown that for a skyrmion lattice in a confining pinned geometry,
application of an ac drive combined with the intrinsic Magnus force produces
circular orbits of skyrmions near the edge of the pin-free channel.
This can generate
an edge current of skyrmions in which the direction of transport
is controlled by the sign of the Magnus force and the
orientation of the edge.
Simultaneously, skyrmions in the bulk of the pin-free channel are localized and follow
closed periodic orbits.
The magnitude of the edge current
depends on the ratio of the Magnus force to the damping term as well as on
the amplitude and frequency of the ac drive.
We identify three dynamic phases:
a localized state in which no edge transport occurs,
a lattice state with edge transport,
and a disordered fluctuating state
that appears for high ac drive amplitude.
In regimes where the edge transport is strong, the transport can extend beyond the
edges and can involve one or two additional rows of skyrmions adjacent to the
edgemost row.
In the overdamped 
limit, the skyrmion orbits become one-dimensional
and the edge transport is lost.
The edge currents are the most pronounced when the
ac drive is parallel to the pin-free channel
since the Magnus force induces motion perpendicular to the driving direction.
A parallel drive pushes the skyrmions against the edges of the confining potential
and induces the circular orbits that are necessary to generate the edge transport.
In contrast, when the ac drive is applied perpendicular to the pin-free channel,
the motion of the skyrmions is more one dimensional, the coupling to the confining
potential is reduced, and the edge transport is diminished.
We show that these results are robust against disorder
and that in some cases the addition of disorder
can induce edge transport.
For lower skyrmion densities,
we observe a skyrmion pump effect
produced by an incommensuration between the number of skyrmions in the row at the
edge of the pin-free channel and the number of skyrmions in the adjacent pin-free row.
The
emergence of  edge currents
should be a generic feature of driven or excited skyrmion lattices
that are subjected to confinement or that contain an interface,
and as an example we
demonstrate that an ac drive can induce edge transport along the domain
boundary separating two different species of skyrmions.

\acknowledgments
This work was supported by the US Department of Energy through
the Los Alamos National Laboratory.  Los Alamos National Laboratory is
operated by Triad National Security, LLC, for the National Nuclear Security
Administration of the U. S. Department of Energy (Contract No. 892333218NCA000001).

\bibliography{mybib}
\end{document}